%% file: CPGutierrez.tex
\documentclass[a4paper,fleqn,usenatbib,useAMS]{mnras}

\usepackage{graphicx}	
\usepackage{amsmath}	
\usepackage{amssymb}	
\usepackage{multicol}        
\usepackage{bm}		
\usepackage{pdflscape}	
\usepackage{graphicx}	
\usepackage{amsmath}	
\usepackage{amssymb}	
\usepackage{multicol}        
\usepackage{bm}		
\usepackage{pdflscape}	
\usepackage{threeparttable,lscape}
\usepackage{natbib}
\usepackage{rotating}
\usepackage{color}
\usepackage{xcolor}
\usepackage{appendix}
\usepackage{multirow, array}
\usepackage{longtable}

\usepackage{etoolbox}
\makeatletter
\patchcmd\@combinedblfloats{\box\@outputbox}{\unvbox\@outputbox}{}{\errmessage{\noexpand patch failed}}
\makeatother

\defcitealias{Dessart14}{D14}
\defcitealias{Anderson16}{A16}
\defcitealias{Taddia16}{T16}

\newcommand{\nodata}{\centering\arraybackslash --} 
\newcommand{\pd}{\ensuremath{P_\mathrm{d}}}
\newcommand{\optd}{\ensuremath{OPT_\mathrm{d}}}
\newcommand{\mstellar}{\ensuremath{M_\mathrm{stellar}}}
\newcommand{\mmax}{\ensuremath{M_V^\mathrm{max}}}
\newcommand{\mbhost}{\ensuremath{M_B^\mathrm{host}}}
\newcommand{\mrhost}{\ensuremath{M_r^\mathrm{host}}}
\newcommand{\pewHb}{\textrm{pEW}\ensuremath{\left(\mathrm{H}\beta\right)}}
\newcommand{\pewHa}{\textrm{pEW}\ensuremath{\left(\mathrm{H}\alpha\right)}}
\newcommand{\pewfe}{\textrm{pEW}\ensuremath{\left(\ion{Fe}{ii}\,5018\right)}}
\newcommand{\pewfeb}{\textrm{pEW}\ensuremath{\left(\ion{Fe}{ii}\,5169\right)}}
\newcommand{\pewBa}{\textrm{pEW}\ensuremath{\left(\ion{Ba}{ii}\,6142\right)}}
\newcommand{\pewSc}{\textrm{pEW}\ensuremath{\left(\ion{Sc}{ii}\,6247\right)}}
\newcommand{\pewNaD}{\textrm{pEW}\ensuremath{\left(\mathrm{Na\,I\,D}\right)}}

\definecolor{yaleblue}{rgb}{0.1,0.3,0.9}
\definecolor{ultramarine}{rgb}{0, 0, 150}
\hypersetup{colorlinks=true, linkcolor=yaleblue, urlcolor=blue, citecolor=yaleblue}

\title[SNe~II in low luminosity host galaxies]{Type II supernovae in low luminosity host galaxies}
\author[C. P. Guti\'errez et al.]
{C.~P. Guti\'errez$^{1}$\thanks{E-mail: C.P.Gutierrez-Avendano@soton.ac.uk},
J.~P. Anderson$^{2}$, M. Sullivan$^{1}$, L. Dessart$^{3}$, S. Gonz\'alez-Gaitan$^{4}$, 
\newauthor L. Galbany$^{5}$, G. Dimitriadis$^{6}$, I. Arcavi$^{7,8}$\footnote{Einstein Fellow}, F. Bufano$^{9}$, 
T.-W. Chen$^{10}$\footnote{Alexander von Humboldt Fellow}, M. Dennefeld$^{11}$,
\newauthor M. Gromadzki$^{12}$, J.~B. Haislip$^{13}$, G. Hosseinzadeh$^{7,8}$, D.~A. Howell$^{7,8}$, C. Inserra$^{1}$,
\newauthor E. Kankare$^{14}$, G. Leloudas$^{15}$, K. Maguire$^{14}$, C. McCully$^{7,8}$; N. Morrell$^{16}$, 
\newauthor F. Olivares E.$^{17,18}$, G. Pignata$^{19,17}$, D.~E. Reichart$^{13}$, T. Reynolds$^{20}$, S.~J. Smartt$^{14}$, 
\newauthor J. Sollerman$^{21}$, F. Taddia$^{21}$, K. Tak\'ats$^{17,19}$, G. Terreran$^{22}$, S. Valenti$^{23}$, D.~R. Young$^{14}$ 
\\
$^{1}$Department of Physics and Astronomy, University of Southampton, Southampton, SO17 1BJ, UK\\
$^{2}$European Southern Observatory, Alonso de C\'ordova 3107, Casilla 19, Santiago, Chile\\
$^{3}$Unidad Mixta Internacional Franco-Chilena de Astronom\'ia (CNRS, UMI 3386), Departamento de Astronom\'ia, Universidad de Chile,\\
Camino El Observatorio 1515, Las Condes, Santiago, Chile \\
$^{4}$CENTRA, Instituto Superior T\'ecnico - Universidade de Lisboa, Portugal\\
$^{5}$PITT PACC, Department of Physics and Astronomy, University of Pittsburgh, Pittsburgh, PA 15260, USA\\
$^{6}$Department of Astronomy and Astrophysics, University of California, Santa Cruz, CA 95064, USA\\
$^{7}$Department of Physics, University of California, Santa Barbara, CA 93106-9530, USA\\
$^{8}$Las Cumbres Observatory, 6740 Cortona Dr Ste 102, Goleta, CA 93117-5575, USA\\
$^{9}$INAF - Osservatorio Astrofisico di Catania, Via Santa Sofia, 78, 95123, Catania, Italy\\
$^{10}$Max-Planck-Institut f{\"u}r Extraterrestrische Physik, Giessenbachstra\ss e 1, 85748, Garching, Germany\\
$^{11}$IAP/CNRS, UMR7095, and Sorbonne Universit\'e 98bis, Boulevard Arago, F-75014 Paris\\
$^{12}$Warsaw University Astronomical Observatory, Al. Ujazdowskie 4, 00-478 Warszawa, Poland\\
$^{13}$University of North Carolina at Chapel Hill, Campus Box 3255, Chapel Hill, NC 27599-3255, USA\\
$^{14}$Astrophysics Research Centre, School of Mathematics and Physics, Queens University Belfast, Belfast BT7 1NN, UK\\
$^{15}$Dark Cosmology Centre, Niels Bohr Institute, University of Copenhagen, Juliane Maries vej 30, 2100 Copenhagen, Denmark\\
$^{16}$Carnegie Observatories, Las Campanas Observatory, Casilla 601, La Serena, Chile\\
$^{17}$Millennium Institute of Astrophysics, Vicuna Mackenna 4860, 7820436 Macul, Santiago, Chile\\
$^{18}$Departamento de Astronom\'ia, Universidad de Chile, Camino el Observatorio 1515, Santiago, Chile\\
$^{19}$Departamento de Ciencias Fisicas, Universidad Andres Bello, Avda. Rep\'ublica 252, Santiago, Chile\\
$^{20}$Tuorla Observatory, Department of Physics and Astronomy, University of Turku, V\"ais\"al\"antie 20, FI-21500 Piikki\"o, Finland \\
$^{21}$The Oskar Klein Centre, Department of Astronomy, AlbaNova, SE-106 91 Stockholm , Sweden\\  
$^{22}$Center for Interdisciplinary Exploration and Research in Astrophysics CIERA, Department of Physics and Astronomy,\\
Northwestern University, Evanston, IL 60208, USA\\
$^{23}$Department of Physics, University of California, Davis, CA 95616, USA\\
}

\date{Accepted XXX. Received YYY; in original form ZZZ}

\pubyear{}

\begin{document}
\label{firstpage}
\pagerange{\pageref{firstpage}--\pageref{lastpage}}
\maketitle

\begin{abstract}
We present an analysis of a new sample of type II core-collapse supernovae (SNe~II) occurring within low-luminosity galaxies, comparing these with a sample
of events in brighter hosts. Our analysis is performed comparing SN~II spectral and photometric parameters and estimating 
the influence of metallicity (inferred from host luminosity differences) on SN~II transient properties. We measure the SN absolute
magnitude at maximum, the light-curve plateau duration, the optically thick duration, and the plateau decline rate in the $V-$band,
together with expansion velocities and pseudo-equivalent-widths
(pEWs) of several absorption lines in the SN spectra. For the SN host galaxies, we estimate the absolute magnitude and the stellar mass,
a proxy for the metallicity
of the host galaxy. SNe~II exploding in low luminosity galaxies display weaker pEWs of \ion{Fe}{ii}\,$\lambda$5018, confirming the 
theoretical prediction that metal lines in SN~II spectra should correlate with metallicity. We also find that SNe~II in low-luminosity
hosts have generally slower declining light curves and display weaker absorption lines. We find no relationship between the
plateau duration or the expansion velocities with SN environment, suggesting that the hydrogen envelope mass and the explosion energy are not correlated with
the metallicity of the host galaxy. This result supports recent predictions that mass-loss for red supergiants is independent of metallicity.
\end{abstract}

\begin{keywords}
supernovae: general -- galaxies: general
\end{keywords}



\section{Introduction}

Type II supernovae (SNe~II) are the terminal explosions of massive ($>8$\,M$_\odot$) stars that have retained a significant 
fraction of their hydrogen envelopes and, hence, have optical spectra that exhibit strong Balmer lines \citep{Minkowski41}. 
Initially, SNe~II were separated into two groups: those with faster declining light curves were classified as SNe IIL, 
while those with a plateau in their light curves (a quasi-constant luminosity for a period of a few months) were classified 
as as SNe IIP \citep{Barbon79}. This distinction has recently been refined with larger samples of events that show a
continuum in their photometric properties \citep[e.g.,][]{Anderson14,Sanders15}. This continuum suggests that 
SNe~II\footnote{Throughout the remainder of the manuscript we use SNe~II to refer to all SNe which would historically have 
been classified as SN IIP or SN IIL. Type IIn, IIb and 87A-like events are excluded from our analysis.} events all come 
from the same progenitor population. The direct identification of the progenitors on pre-explosion images \citep{Smartt15} has 
shown this population to be red supergiants (RSGs).

However, a significant diversity in the properties of SNe~II is observed. Large samples of events have begun to provide
some understanding of this diversity \citep[e.g.,][]{Arcavi12, Anderson14, Faran14a, Faran14b, 
Pejcha15, Pejcha15a, Gonzalez15, Valenti16, Galbany16, Rubin16, Gutierrez17a, Gutierrez17b}. It has been found that 
SNe~II with faster decline rates exhibit a shorter plateau duration \citep[][]{Anderson14,Valenti16,Gutierrez17b}. 
In addition, they are  brighter at 50 days from explosion and during the radioactive tail phase, and have broader
spectral absorption features \citep[e.g.,][]{Hamuy03, Pastorello03, Faran14b, Pejcha15, Pejcha15a,Gutierrez17b}.
These results suggest that the diversity is produced by differences in the progenitor and the explosion mechanisms
(e.g., the amount of the hydrogen envelope mass, the explosion energy, the radius, metallicity, and mass loss). 

SNe~II have been proposed as environmental metallicity probes. \citet[][hereafter D14]{Dessart13a,Dessart14} 
presented SN~II spectral models produced from progenitors with different metallicities: 0.1, 0.4, 1 and 2 times solar 
metallicity (Z$_\odot$). With these models, they predict that the strength of the metal lines during the recombination 
phase should be related to the metallicity of the SN progenitor. They also note a lack of SNe~II at metallicities below
0.4\,Z$_\odot$, supporting the results of \citet{Stoll13}. Using a sample of 119 SNe~II and gas-phase metallicity 
estimates derived from emission line measurements, \citet[][hereafter A16]{Anderson16} confirmed this prediction by
showing a correlation between gas-phase metallicity and SN~II pseudo-equivalent-widths (pEWs) of the \ion{Fe}{ii} $\lambda$5018
line (\pewfe): SNe~II exploding in higher-metallicity galaxies have stronger iron lines in their spectra compared to 
those in lower-metallicity environments. However, no trend was seen with any other SN~II property, and \citetalias{Anderson16}
therefore concluded that progenitor metallicity likely plays only a minor role in driving SN~II diversity. However, they noted 
that the range in host-galaxy luminosity sampled was not particularly large ($-18\gtrsim$\mbhost$\gtrsim-22$). 

This restricted range in host-galaxy luminosity has now been overcome with new surveys that scan large areas of the sky 
without preference to the location of bright, nearby galaxies. Such searches have found new trends with respect to the ratios
of different SN types as a function of galaxy luminosity \citep[see][]{Arcavi10}, but detailed studies of the properties
of SNe~II located in low-luminosity hosts are still generally lacking.

A recent exception is \citet[][hereafter T16]{Taddia16}, who analysed a further 39 SNe~II taken from the (intermediate) 
Palomar Transient Factory sample of \citet{Rubin16}, extending the sample to lower-metallicity with 18 events. These events
showed smaller \pewfe, with a weak correlation between the inferred host-galaxy metallicity (using an average luminosity--colour--metallicity relationship) 
and \pewfe. In addition, \citetalias{Taddia16} showed that SNe~II with brighter peak magnitudes tend to occur at lower metallicity.

This paper builds on these earlier studies, analysing a larger number of SNe~II in low-luminosity hosts and their properties. 
Of particular interest is the duration of the `plateau' phase in SN~II light curves (\pd). This has long been linked to the mass
of the hydrogen-rich envelope of the progenitor star at the time of explosion \citep[e.g.,][and recent discussion in 
\citealt{Gutierrez17b}]{Popov93}, as \pd\ is believed to be directly related to the time the hydrogen envelope takes to fully
recombine. The hydrogen envelope mass is itself related to the mass-loss suffered by the progenitor star during its lifetime
and the amount of hydrogen that has been fused into higher mass elements in the core.
Given the metallicity dependence of mass-loss for hot single stars \citep[e.g.][]{Vink01,Mokiem07}, most stellar models
predict a strong dependence of core-collapse SN type on progenitor metallicity \citep[e.g.][]{Heger03,Chieffi13}. This 
metallicity dependence of mass loss is also presumed to  affect the hydrogen envelope mass retained by SN~II progenitors, 
leading to a predicted dependence of the \pd\ on progenitor metallicity.

It should be noted, however, that the majority of the mass loss suffered by a SN~II progenitor will happen during the RSG 
phase, and currently there is no strong observational evidence that the strength of RSG mass loss correlates with metallicity.
A recent study by \citet{Chun17} showed that metallicity dependence of mass loss is only apparent when the Schwarzschild 
criterion for convection is employed; using the Ledoux criterion, the envelope mass at the epoch
of explosion for SNe~II was almost independent of progenitor metallicity. This result is supported by \citet{Goldman17}, who
found no metallicity dependence of mass-loss for RSGs in the Large Magellanic Cloud (LMC) between half and twice solar metallicity. 
Within this context, observations of SNe~II arising from a large range of environmental (and therefore progenitor) metallicity
could be highly constraining for stellar models.

This paper presents 30 new SNe II in low-luminosity host galaxies. Using measurements of the SN photometric and spectroscopic
properties, our aim is to further test the validity of using SNe~II as metallicity indicators, and 
to constrain massive star models and SN~II progenitors by searching for correlations of SN~II parameters with host 
galaxy properties. The paper is organised as follows. In Section~\ref{sample} we describe our sample and observations.
The measurements are presented in Section~\ref{measurements}, while the results and discussion are presented in 
Section~\ref{results} and Section~\ref{discussion}, respectively. We conclude in Section~\ref{conclusions}. Throughout,
we assume a flat $\Lambda$CDM universe, with a Hubble constant of $H_0=70$\,km\,s$^{-1}$\,Mpc$^{-1}$, and
$\Omega_\mathrm{m}=$0.3.

\begin{figure*}
\centering
\includegraphics[width=14cm]{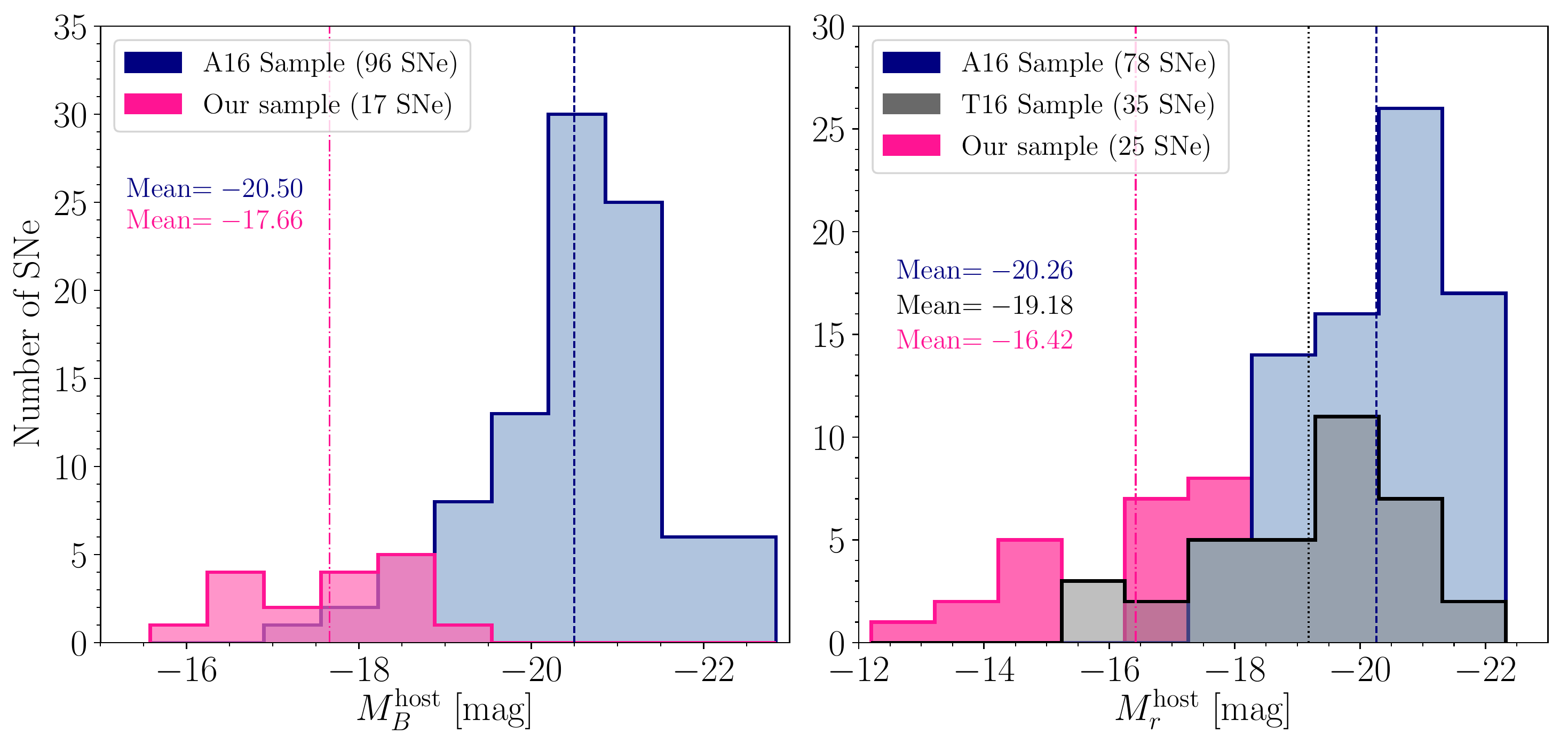}
\caption{Distribution of the host-galaxy absolute magnitudes for our sample (magenta), and for the \citetalias{Anderson16} 
(blue) and \citetalias{Taddia16} (black) samples. The left panel shows the distribution of \mbhost, and the right panel
the distribution of \mrhost. The vertical lines indicate the mean for each sample (dashed: \citetalias{Anderson16}, dotted: 
\citetalias{Taddia16}, and dot-dashed: our sample). Note that not all of our 30 SNe~II have host information in both filters
(see Section~\ref{sec:hostgalaxydata} for further details).}
\label{histom}
\end{figure*}

\section{Data Sample and observations}
\label{sample}

In this paper, we present a new sample of 30 SNe~II in low-luminosity host galaxies, and combine with a further 108 events
sampling a range of host galaxy luminosities and taken from the literature (\citetalias{Anderson16}, \citealt{Gutierrez17a}, 
and references therein). Our new SNe~II were selected so that they were i) located in galaxies with an absolute magnitude in 
the $B$-band (\mbhost) of $\mbhost\gtrsim-18.5$\,mag, or were apparently hostless, and ii) younger than 20 days post explosion. This absolute
magnitude limit was chosen as it is around the brightness of the LMC, and because very few SN~II were
observed in the sample used by \citetalias{Dessart14} with implied metallicities below that of the LMC. In this section
we present the data on the new sample of events, including their optical photometry and spectroscopy, and the data on their host 
galaxies.

\subsection{SN sample}
\label{sec:snobservations}

Observations for the new low-luminosity host sample of 30 objects are drawn from several sources. We took data on SNe~II from the Public ESO 
Spectroscopic Survey for Transient
Objects \citep[PESSTO;][and the extended PESSTO, ePESSTO]{Smartt15a}, which specifically targeted SNe~II in faint host galaxies,
and from the Las Cumbres Observatory
\citep[LCO;][]{Brown13} Supernova Key Project and the LCO Global Supernova Project. These events were generally observed photometrically
with a three day cadence, and with spectral observations every few weeks. Finally, one additional SN~II was detected and observed in
2009 by the CHilean Automated Supernova sEarch \citep[CHASE;][]{Pignata09}.
Our comparison literature SNe~II are taken from the various samples observed between 1986 and 2009 and compiled in \citet{Anderson14}. 
These comprise: the Cerro Tololo Supernova Survey (CTSS),
the Cal\'an/Tololo survey (CT, \citealt{Hamuy93}), the Supernova Optical and Infrared Survey (SOIRS),
the Carnegie Type II Supernova Survey (CATS) and the Carnegie Supernova Project (CSP-I, \citealt{Hamuy06}).
Table~\ref{table_info} gives the SN host galaxy information: 
recession velocity, \mbhost\ (used for the sample selection) and the reddening
due to dust in our Galaxy ($E(B-V)_\mathrm{MW}$), as well as details of the discovery date and explosion epoch of each SN.

In Fig.~\ref{histom} we present the \mbhost\ and \mrhost\ distributions of our sample in comparison to \citetalias{Anderson16}
and \citetalias{Taddia16}. Our 30 new SNe extend the host-galaxy luminosity distribution to fainter hosts, with a mean 
$\mrhost=-16.42\pm0.39$\,mag (cf. $\mrhost=-20.26\pm0.14$\,mag in \citetalias{Anderson16}, and $\mrhost=-19.18\pm0.27$\,mag in \citetalias{Taddia16}).

\subsubsection{Photometry}
\label{sec:photometry}

Optical photometry was acquired for 29 of the 30 new low-luminosity host SNe~II. The light curves of 24 SNe~II were obtained by LCO, either as part of 
PESSTO, or as part of the LCO key projects, and reduced following the prescriptions described by \citet{Firth15}. 21 of these SNe
were observed in $BVgri$ filters, while the remaining four in $gri$. SN~2017vp was observed in the $g'r'i'z'JHK$ bands with the Gamma-Ray Burst
Optical/Near-Infrared Detector (GROND; \citealt{Greiner08}), at the 2.2-m MPG telescope at the European Southern Observatory 
(ESO) La Silla Observatory in Chile. The images were reduced with the GROND pipeline \citep{Kruhler08}, which applies de-bias
and flat-field corrections, stacks images and provides an astrometric calibration. SN~2015bs was observed in $BVri$ using 
the Swope telescope at the Las Campanas Observatory. The reduction procedure is presented in \citet{Anderson18}.
SN~2009lq, ASASSN-14kp and SN~2014cw were observed in $BVRI$ with the PROMPT telescopes located at Cerro Tololo Interamerican
Observatory. SN~2009lq was, in addition, observed with the TRAPPIST telescope at La Silla. The reduction of the images of 
these three SNe were performed following standard procedures (including  bias, dark, and flat-field corrections), and calibrated 
using observations of standard-star fields \citep{Landolt92, Landolt07, Smith02}. ASASSN-15rp has no photometric information. 
Details of the literature sample can be found in \citet{Anderson14}.

Table~\ref{obs} presents a summary of all the observations obtained for the new low-luminosity host
SNe~II in our sample. The photometric data for SN~2016X and SN~2015bs are presented by \citet{Huang18} and \citet{Anderson18},
respectively, while the data for ASASSN-14dq and SN~2015W were published in \citet{Valenti16}. Published data 
have been re-reduced and calibrated here for homogeneity within sample. Data (both photometric and spectroscopic) for ASASSN-15oz
will be presented in a separate study by Bostroem (in preparation), for SN~2016B in Rui et al. (in preparation), for SN~2016blz in
Johansson et al. (in preparation), for SN~2016dbm in Terreran et al. (in preparation), for SN~2016egz in Hiramatsu (in preparation), 
for SN~2016enp in Reynolds et al. (in preparation), and for SN~2016gsd in Reynolds et al. (in preparation). Photometry and 
spectroscopy for the remaining unpublished SNe in the sample will be presented in a future data paper.

Throughout the paper, all magnitudes in our sample have been corrections for Milky Way extinction taken from \citet{Schlafly11}. 
K-corrections were not applied because at low redshift the results are not affected. 
See \citet{Anderson14} for more details.

\subsubsection{Spectroscopy}
\label{sec:spectroscopy}

In this paper, we make use of spectroscopic measurements at (or close to) 50 days after the SN explosion. These spectra come from 
a range of difference sources. Details of the instruments used for the spectral observations of the new sample are listed
in Table~\ref{obs}, and all spectra from which we make the $+$50\,d measurements in this paper are available through the 
WISeREP\footnote{\url{http://wiserep.weizmann.ac.il/home}} archive \citep{Yaron12}. Spectroscopic data for SN~2015bs is presented
in \citet{Anderson18}, and for SN~2016X in \citet{Huang18}.
The PESSTO spectra up to May 2016 can also be retrieved from the ESO Science Archive Facility as Phase 3 data 
products\footnote{See \url{http://www.pessto.org} for access instructions.}.

Spectroscopic observations were performed with the ESO Faint Object Spectrograph and Camera \citep[EFOSC,][]{Buzzoni84}
at the 3.5-m ESO New Technology Telescope (NTT), and the twin FLOYDS spectrographs on the Faulkes Telescope South (FTS) 
and the Faulkes Telescope North (FTN). The NTT spectra were reduced using the PESSTO pipeline \citep{Smartt15a}, while 
the FLOYDS data were reduced using the PyRAF-based \texttt{floydsspec} pipeline. Data for SN\,2014cw were acquired with 
the Low Resolution Spectrograph (LRS) at the 3.6-m Telescopio Nazionale Galileo (TNG), the Ohio State Multi-Object Spectrograph 
(OSMOS) at the 2.4-m Hiltner Telescope, and the Goodman Spectrograph at the SOAR 4.1-m telescope. For SN\,2009lq, two spectra 
were obtained with the Wide Field CCD Camera (WFCCD) at the 2.5-m du Pont Telescope and the Low Dispersion Survey Spectrograph
(LDSS3) on the Magellan Clay 6.5-m telescope at Las Campanas Observatory. For ASASSN-14kp one spectrum was obtained with WFCCD. 
The reductions for LRS, TNG, OSMOS, Goodman, WFCCD and LDSS3 spectra were performed using the standard routines (bias subtraction,
flat-field correction, 1D extraction, and wavelength calibration). Details on the literature spectra are
presented by \citet{Gutierrez17a}.

\subsection{Host galaxy data}
\label{sec:hostgalaxydata}

Photometry for the SN host galaxies in the $ugriz$ filters were collected from the Sloan Digital Sky Survey (SDSS) Data Release
13\footnote{\url{http://skyserver.sdss.org/dr13/en/home.aspx}} \citep[][]{Albareti17} and the
Pan-STARRS1\footnote{\url{https://panstarrs.stsci.edu/}} \citep{Flewelling16,Chambers16} data archive. In addition, $B-$band 
photometry (used in our initial SN selection  for spectroscopic and photometric follow-up) was collected from the 
HyperLEDA\footnote{\url{http://leda.univ-lyon1.fr/}} \citep{Makarov14} database. For SN\,2015bs, \mrhost\ 
was obtained from \citet{Anderson18}, while for SN~2009lq, ASASSN-15oz, ASASSN-15rp, and SN~2016drl, where no host 
galaxy is visible, we determine an upper limit on the luminosity of the host using a circular aperture centered on the SN location.	
The host  galaxy magnitudes can be found in Table~\ref{mag}.

\section{Measurements}
\label{measurements}

We now turn to the measurements that we make on the SN photometry and spectra, and the host galaxies.

\subsection{SN measurements}
\label{sec:sn-measurements}

\begin{figure}
\centering
\includegraphics[width=\columnwidth]{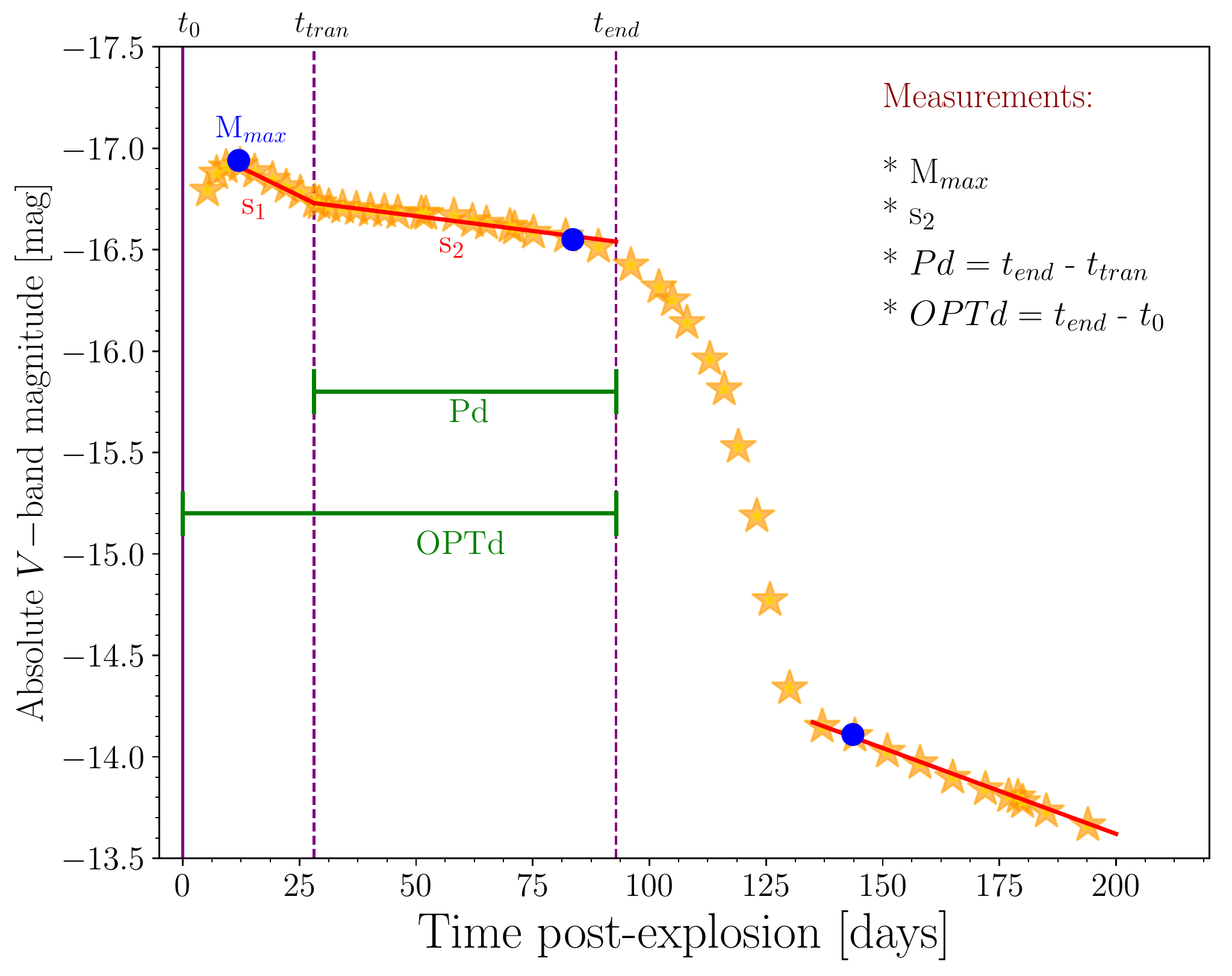}
\includegraphics[width=\columnwidth]{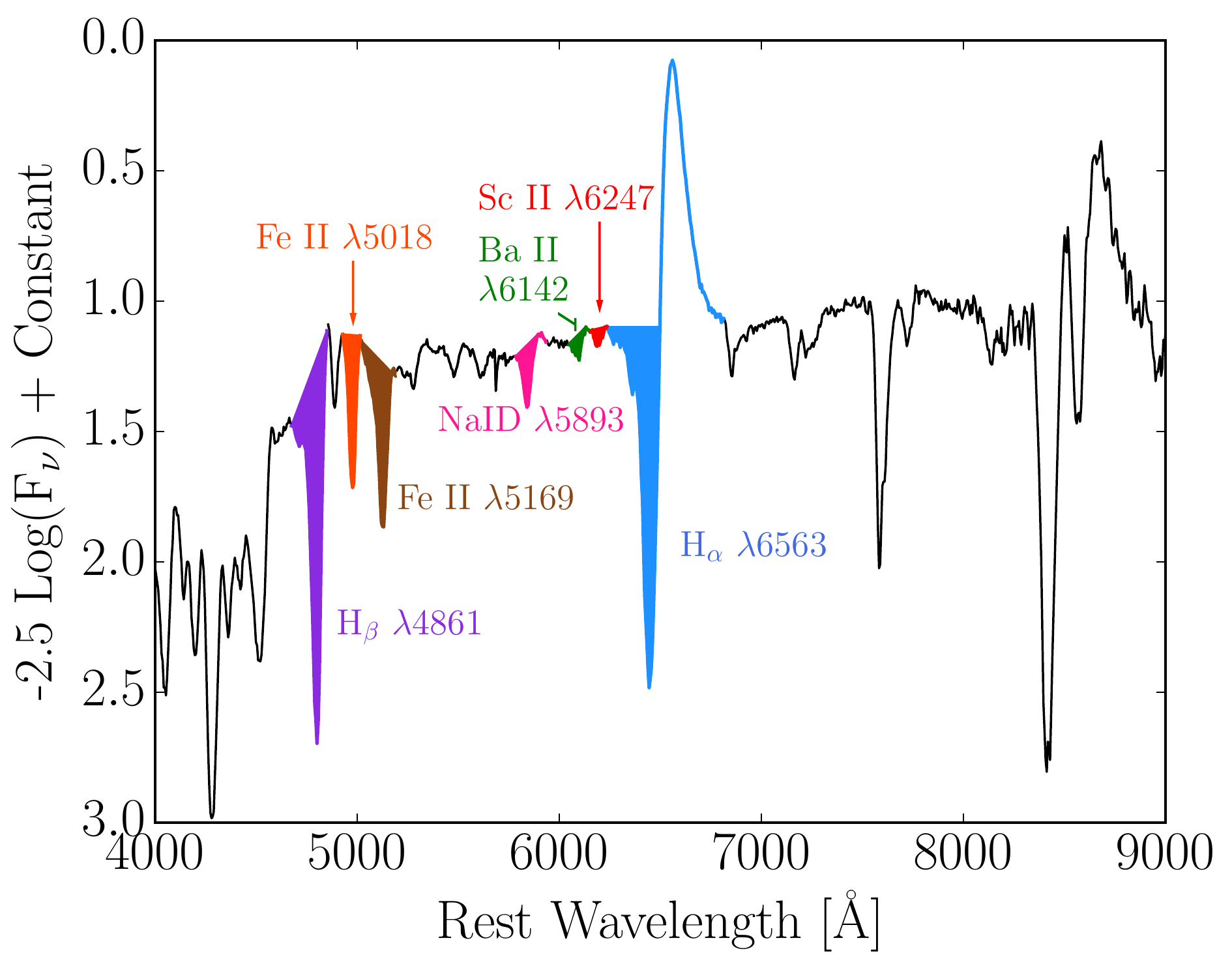}
\caption{Definition of the measurements for our sample. The upper panel shows the light-curve parameters measured for each SN 
in the $V$-band. The lower panel shows the absorption lines for which we measure pEWs, shown in the photospheric +41\,d spectrum of SN\,2016aqf.}
\label{examples}
\end{figure}

We measure several photometric and spectroscopic parameters in our SN~II sample. These spectral and photometric 
measurements were performed following the prescriptions presented in \citet{Anderson14} and \citet{Gutierrez17a}. From
the photometry, we measure in the $V$-band the magnitude at maximum light($M_\mathrm{max}$), 
the plateau decline rate ($s_2$), the optically-thick duration phase (\optd) and the plateau duration (\pd). 
Here, we define \pd\ as in \citet{Anderson14}: the duration from the inflection point (t$_\mathrm{tran}$) between the initial decline ($s_1$)
and the plateau decline ($s_2$), until the end of the plateau (t$_\mathrm{end}$). We measured \optd\ between the explosion (t$_0$) and the 
end of the plateau (t$_\mathrm{end}$), while $s_2$ was measured by fitting a straight line during the plateau phase. Given that \pd\ is estimated 
over the interval [t$_\mathrm{tran}$, t$_\mathrm{end}$], the number of SNe with available estimates is significantly smaller compared to \optd\ or s$_2$.
This is because a better photometric coverage is necessary to calculate the t$_\mathrm{tran}$ and the t$_\mathrm{end}$, compared to $s_2$.
Fig.~\ref{examples} presents an example light curve indicating these parameters.

From the spectra, we measure the expansion velocities and the pEWs for 7 lines: H$\alpha$, H$\beta$, \ion{Fe}{ii}\;$\lambda$5018,
\ion{Fe}{ii}\;$\lambda$5169, \ion{Na}{i}\;D, \ion{Ba}{ii}\;$\lambda$6142 and \ion{Sc}{ii}\;$\lambda$6247.
We estimated the expansion velocities using the relativistic Doppler equation and the rest wavelength of each line.
Further details are presented in \citet{Gutierrez17a}.
The pEWs were measured by tracing a straight line across the absorption feature to mimic the continuum flux, as shown in the
lower panel of  Fig.~\ref{examples}. 
Note that in this paper, \ion{Na}{i}\,D refers to the Doppler-broadened \ion{Na}{i}\,D line forming in the SN ejecta, and not the narrow line
caused by any surrounding circumstellar material or interstellar material along the line-of-sight in the SN host. The spectral parameters are 
measured at (or interpolated to) 50 days from explosion ($+50$\,d). We estimate the explosion epochs using the prescriptions described in 
\citet{Gutierrez17a}, and the epochs are listed in Table~\ref{table_info} together with the technique employed to determine 
them. Details about the explosion epoch estimates for the literature events can be found in \citet{Gutierrez17a}. The final spectroscopic
and photometric properties are presented in Table~\ref{table_measure}.

\subsection{Galaxy measurements}
\label{sec:galaxy-measurements}

We characterise the global properties of the SN~II hosts using the stellar mass (\mstellar). We use a custom galaxy SED fitting code, 
following \citet{Sullivan10}.
The code is similar to \textsc{z-peg} \citep{LeBorgne02}, with an expanded set of templates based on the galaxy spectral 
synthesis code \textsc{p\'{e}gase}.2 \citep{Fioc97}. Specifically, the code uses a set of 15 exponentially declining star
formation histories (SFHs), each with 125 age steps, assuming a \citet{Salpeter55} initial mass function (IMF). The internal 
\textsc{p\'{e}gase}.2 dust prescription is not included in the templates, but instead extinction is included as a foreground 
dust screen with $E(B-V)$=0 to 0.3\,mag in steps 0.05\,mag. The code determines the best-fitting SED model by minimisation of
the $\chi^{2}$ using data in the $uBgriz$ filters. The redshift in the fit is fixed to the value of the SN redshift.

The \mstellar\ of the host galaxy is calculated by integrating the star formation history of the best-fitting SED model, subtracting the 
mass of the stars that have died. The \mstellar\ uncertainties are calculated by performing a Monte Carlo simulation on the observed galaxy 
fluxes according to the uncertainties retrieved from SDSS. Table~\ref{mag} gives the final information on the host galaxies.

\section{Results}
\label{results}

In this section, we compare the SNe~II in low-luminosity galaxies with those located in high-luminosity hosts, searching for  
significant differences in SN properties between the two samples. All spectral comparisons are performed at +50\,d.

\subsection{SNe~II in fainter galaxies vs. SNe~II in brighter galaxies}

We split our full sample into two: 35 SNe~II located within `faint' host galaxies, defined as $M_r\gtrsim-18.5$\,mag (or 
$M_B\gtrsim-18.5$\,mag, where no $r$-band data is available), and 103 SNe~II in `bright' host galaxies, defined as 
$M_r<-18.5$\,mag (or $M_B<-18.5$\,mag). This absolute magnitude limit is around the brightness of the well-studied LMC. This was 
chosen as the separation limit because very few SN~II were observed in the sample used by \citetalias{Dessart14} with inferred 
metallicities below that of the LMC, hence we investigate whether SNe~II dimmer than this limit are distinct from those in brighter hosts.
Details about the characteristics of low- and high-luminosity groups are presented in Table~\ref{table_summary}.

\begin{figure*}
\centering
\includegraphics[width=13cm]{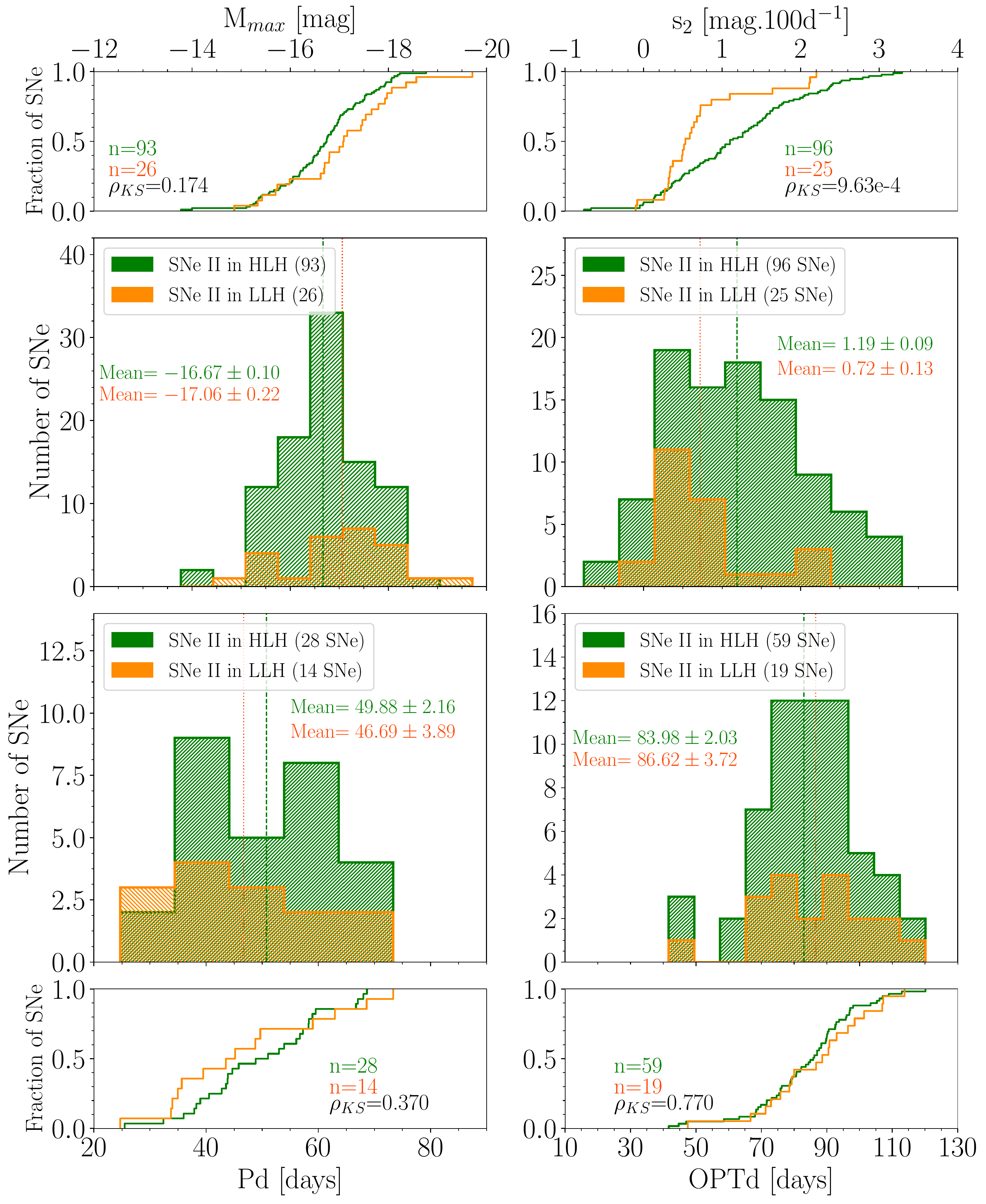}
\caption{The distribution of the SN~II light curve properties \mmax\ (top left), $s_2$ (top right), \pd\ (bottom left) and \optd\ (bottom right) 
in low-luminosity hosts (LLH; orange) and high-luminosity hosts (HLH; green). The upper row shows the histograms of each property, and the lower row 
shows the cumulative distributions. The vertical lines indicate the mean for each distribution. In the lower panels, the number of
events ($n$) and the $\rho_\mathrm{KS}$ are given.}
\label{fig:histophoto}
\end{figure*}

\subsubsection{SN photometric properties}

The photometric properties for our sample are shown in Fig.~\ref{fig:histophoto}. Our SNe~II cover a wide range of observed 
properties, with no clear distinction between those in low- and high-luminosity hosts for \mmax, \pd\ and \optd. However, the mean 
$s_2$ in high-luminosity hosts is $1.19\pm0.09$\,mag\,100d$^{-1}$, compared to $0.72\pm0.13$\,mag\,100d$^{-1}$ in low-luminosity 
hosts. This suggests that SNe~II in lower luminosity hosts have a shallower decline during the recombination phase. 

Fig.~\ref{fig:histophoto} also shows the cumulative distributions for the three photometric parameters. Throughout this paper, 
we compare such cumulative distributions of SN parameters for events in low- and high-luminosity host galaxies. For this we use the 
Kolmogorov-Smirnov (KS) test, a nonparametric test that compares the cumulative distributions of two data sets under the null hypothesis
that both groups are sampled from populations with identical distributions. We reject this null hypothesis when $\rho_\mathrm{KS}<0.01$. 
In this case (Fig.~\ref{fig:histophoto}), the KS test shows that we can reject the null hypothesis that both $s_2$ samples (in high- and
low-luminosity hosts) are drawn from parent populations with the same distributions. The mean values and corresponding KS test results are 
in Table~\ref{table_values}.

\subsubsection{SN spectral properties}

\begin{figure}
\centering
\includegraphics[width=\columnwidth]{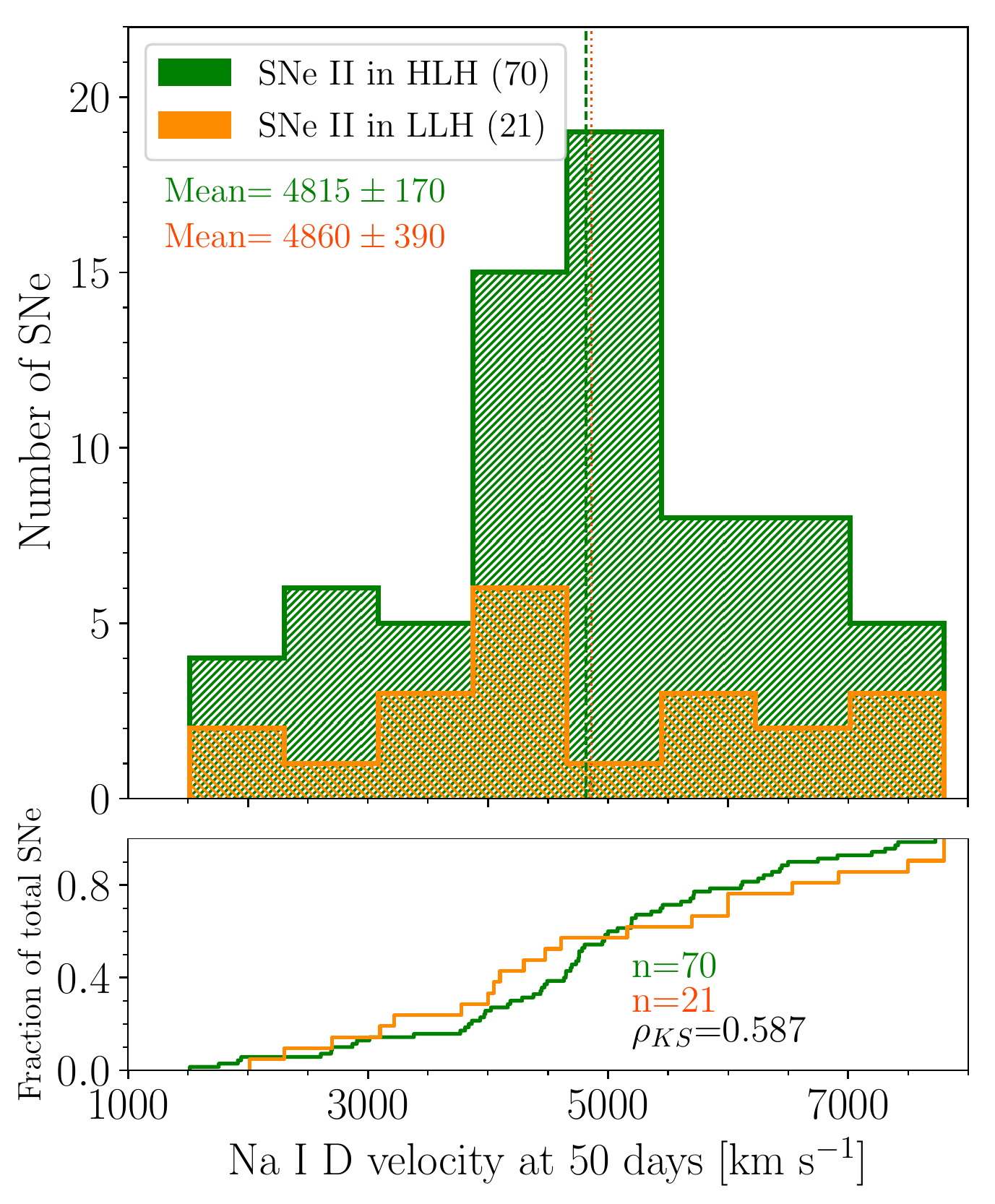}
\caption{The distribution of \ion{Na}{i}\,D velocity at 50 days from explosion (upper panel) and the cumulative distribution 
(lower panel). SNe~II in low-luminosity hosts are in orange, and those in high-luminosity hosts in green. The vertical lines 
indicate the mean for each distribution. In the cumulative distribution panel, the number of SNe and the $\rho_\mathrm{KS}$ are shown.}
\label{fig:vel}
\end{figure}

We see no significant differences in the velocities of SNe~II located in low- and high-luminosity hosts. An example 
is shown in Fig.~\ref{fig:vel}, showing the expansion velocity for \ion{Na}{i}\,D. Only small differences of 
$\sim50$\,km\,s$^{-1}$ are observed. The figure also shows that both groups have a similar velocity distribution, with a large
range of velocity values. Similar behaviour is  observed for the rest of the SN velocities, suggesting that the SN explosion
energy is not significantly affected by metallicity. A complete set of measurements can be found in Table~\ref{table_values}.

\begin{figure*}
\centering
\includegraphics[width=13cm]{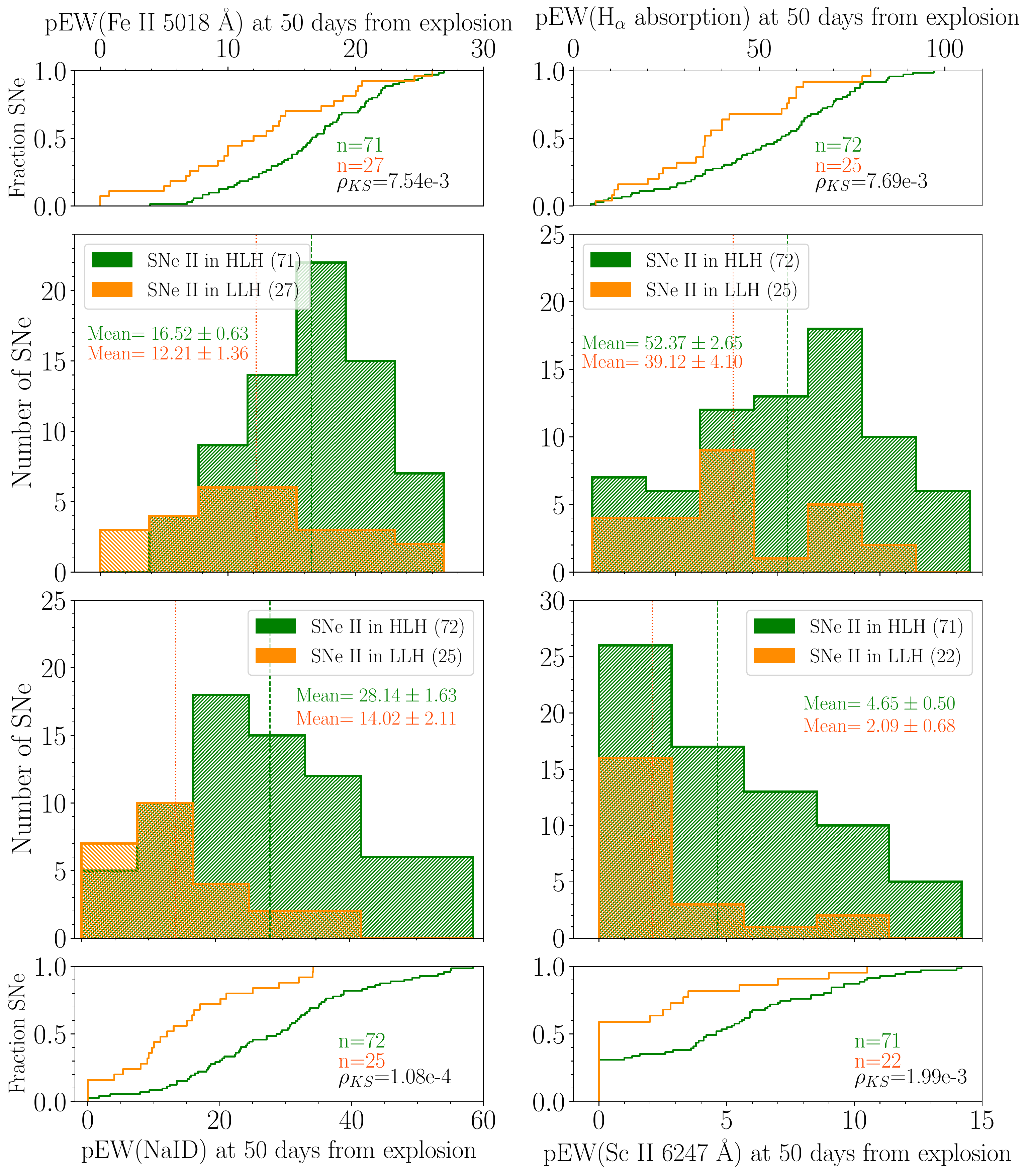}
\caption{Distribution of the pEW absorption measurements for SNe~II in the low-luminosity host sample (orange) in comparison to 
those in the high-luminosity host sample (green). The vertical lines indicate the mean for each distribution: \pewfe, \pewHa, 
\pewNaD\ and \pewSc. The upper and lower panels show the cumulative distributions of each pEW in the two host categories:
\pewfe\ (upper left), \pewHa\ (upper right), \pewNaD\ (lower left) and \pewSc\ (lower right). Also shown are the number of events
($n$), and $\rho_\mathrm{KS}$.}
\label{fig:ew}
\end{figure*}

Fig.~\ref{fig:ew} shows the pEW distributions for \ion{Fe}{ii}\;$\lambda$5018, H$\alpha$ absorption, \ion{Na}{i}\,D and 
\ion{Sc}{ii}\;$\lambda$6247 for the high- and low-luminosity host samples. In general,
the pEWs for SNe~II in low-luminosity hosts are shifted to smaller values. The result obtained for \pewfe\ is consistent with 
the prediction of \citetalias{Dessart14}, who found that the strength of the metal lines increases with increasing progenitor 
metallicity; SNe~II with metal-poor progenitors should have weaker iron lines. Using host luminosity as a proxy for host metallicity 
confirms this prediction observationally. The \ion{Na}{i}\,D line shows a similar behaviour, but,
for the H$\alpha$ line, the opposite to that predicted by \citetalias{Dessart14} is observed, where higher metallicity models
show weaker H$\alpha$ absorption. The comparison of the pEWs of H$\beta$, \ion{Fe}{ii}\;$\lambda$5169 and \ion{Ba}{ii}\;$\lambda$6142
for SNe~II for low- and high-luminosity hosts showed similar results to those in Fig.~\ref{fig:ew}.

The cumulative distributions for the pEW differences are also shown in Fig.~\ref{fig:ew}. KS tests reject the null hypothesis
that the distributions of the pEWs in low- and high-luminosity hosts are drawn from populations with the same distribution for all  four lines. 
The differences are most significant for \ion{Na}{i}\,D and \ion{Sc}{ii}\;$\lambda$6247, with $\rho_\mathrm{KS}$ less than $\sim0.0001$ and 
$\sim0.002$, respectively. The mean pEWs obtained for all 7 lines in the low- and high-luminosity
hosts, and the corresponding KS test values, are presented in Table~\ref{table_values}.

\begin{figure*}
\centering
\includegraphics[width=\textwidth]{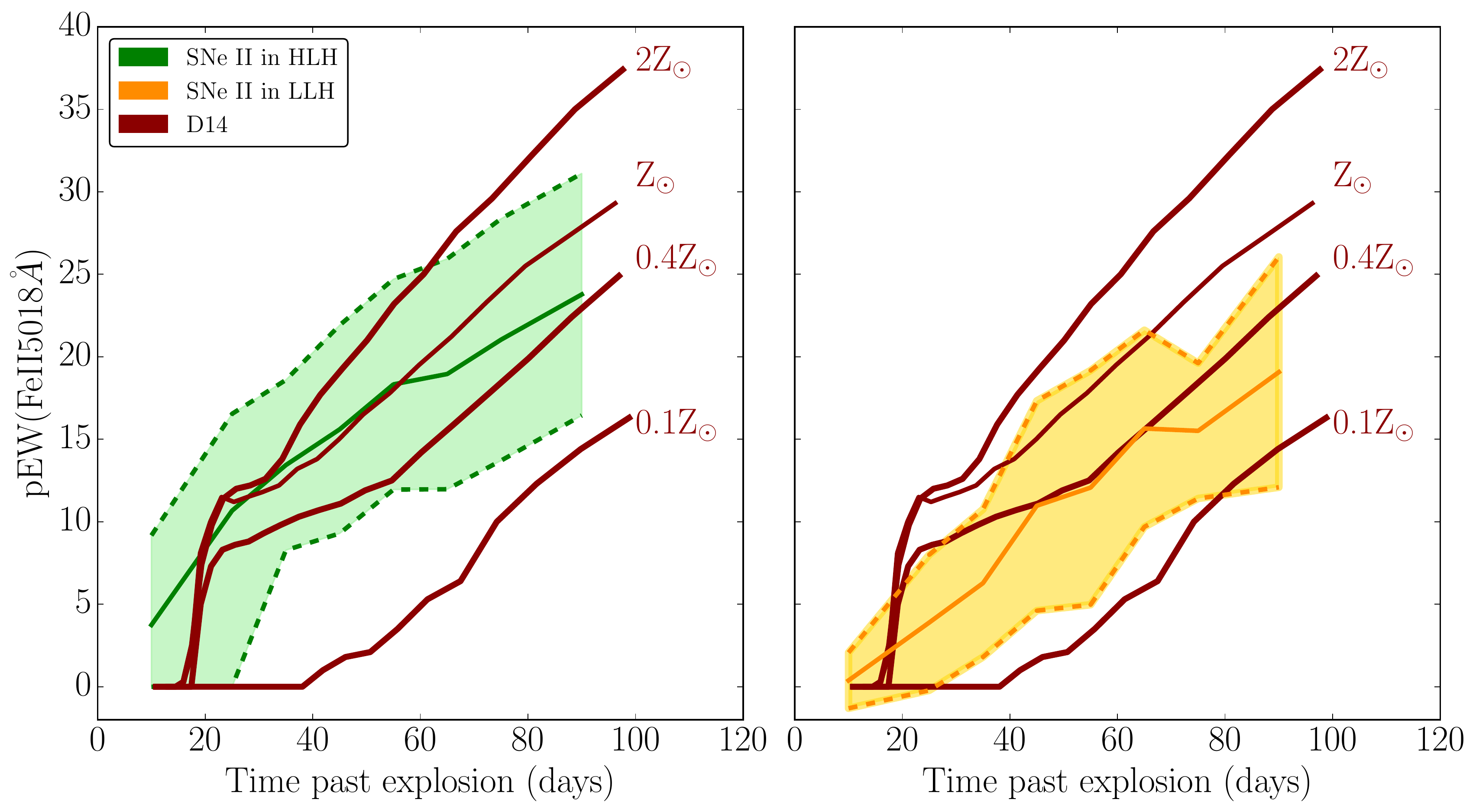}
\caption{Temporal evolution of \pewfe\ from explosion to +90\,d, in both the data and models at four different metallicities. 
The observational data are binned in time. Left: Comparison between the SNe~II in high-luminosity hosts. The green solid line
represent the mean of the pEW within each time bin, while the dashed green lines indicate the standard deviation. \textit{Right:} 
The comparison with SNe~II in low-luminosity hosts, with the mean of the pEW orange solid line and the standard deviation in
dashed orange lines. In both panels, the brown solid lines represent the \citetalias{Dessart14} models as labelled.}
\label{fig:compa}
\end{figure*}

Fig.~\ref{fig:compa} shows the temporal evolution of \pewfe, compared to the synthetic spectra at four different metallicities
\citepalias[see][]{Dessart14}. \pewfe\ for SNe~II in high-luminosity hosts lies around the model with solar metallicity (Z$_\odot$),
while SNe II in low-luminosity hosts are closer to the 0.4\,Z$_\odot$ model. This is again consistent with the low-luminosity host
group sampling SN~II progenitors of lower metallicity than the high-luminosity group.

Fig.~\ref{fig:compa} also shows that the \pewfe\ for SNe~II in high-luminosity hosts evolves faster at early phases. To quantify
this, we measure the rate of change of the pEW ($\Delta$pEW) over the intervals $[+10,+35]$\,d, $[+35,+55]$\,d and $[+55,90]$\,d. 
For low-luminosity hosts, $\Delta$pEW is $0.24\pm0.04$, $0.29\pm0.07$ and $0.20\pm0.08$\,\AA\,d$^{-1}$, while in high-luminosity
hosts the values are $0.39\pm0.03$, $0.25\pm0.05$ and $0.16\pm0.04$\,\AA\,d$^{-1}$: at early phases, the pEW evolves $\sim40$
per cent faster in SNe~II in high-luminosity hosts than in low-luminosity host, but becomes consistent at later phases.

In the models, for the 0.4\,Z$_\odot$ model, the $\Delta$pEW is $0.42\pm0.05$, $0.13\pm0.05$ and $0.30\pm0.0.06$\,\AA\,d$^{-1}$,
while for the Z$_\odot$ model they are $0.53\pm0.03$, $0.27\pm0.04$ and $0.29\pm0.05$\,\AA\,d$^{-1}$.
In the first two intervals, the pEW at Z$_\odot$ evolves faster than the pEW at 0.4Z$_\odot$, but in the last interval, they
are almost the same. Comparing the models with the observed SNe, the Z$_\odot$ model displays a faster evolution
at all epochs in comparison with observed SNe~II, however for the model at lower metallicity, the evolution is faster only 
in the first interval.

\subsection{Correlations between SN~II properties and host-galaxy parameters}
\label{correl}

We next investigate any dependence of SN~II diversity on environmental parameters, specifically the absolute magnitudes and 
stellar masses of the host galaxies. In particular, we examine \pewfe\ and \pewNaD\ for our 
SNe~II in the context of the SN host galaxies. We use \pewfe\ as the models show it is mainly affected by the metallicity
of the progenitor, and we use \pewNaD\ due to the significance of the relations seen in Fig.~\ref{fig:ew} and presented
in Table~\ref{table_values}. 

\begin{figure*}
\centering
\includegraphics[width=\textwidth]{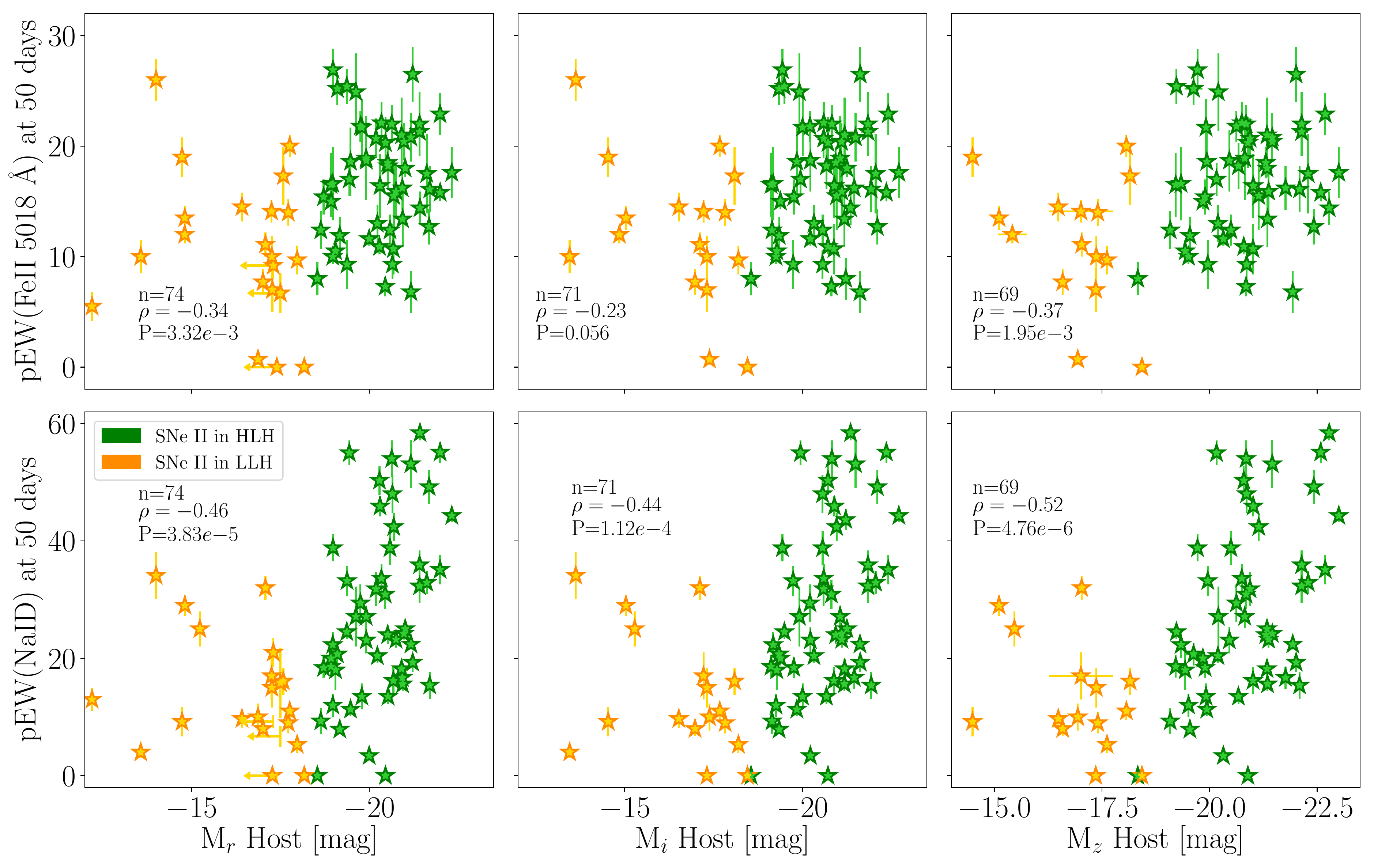}
\caption{Correlations between \pewfe\ (upper panels) and \pewNaD\ (lower panels), and the galaxy absolute magnitudes: $M_r$
(left), $M_i$ (centre) and $M_z$ (right). Green stars represent SNe in high-luminosity hosts, while yellow stars are SNe~II
in low luminosity hosts. In each panel, $n$ is the number of events, $\rho$ is the Pearson's correlation coefficient, and 
$P$ is the probability of detecting a correlation by chance. }
\label{fig:ewhost}
\end{figure*}

Fig.~\ref{fig:ewhost} shows the relation between the host-galaxy absolute magnitudes in different filters ($riz$), and \pewfe\ 
and \pewNaD. In general, SNe~II in lower luminosity hosts have smaller pEW values. Using the Pearson correlation test,  
we find weak anti-correlations for \pewfe\ of $\rho=-0.34$, $-0.23$ and $-0.37$, and moderate anti-correlations for 
\pewNaD: $\rho=-0.46$, $-0.44$, and $-0.52$. The strongest correlations are observed in the $z$-band. 

\begin{figure}
\centering
\includegraphics[width=7.7cm]{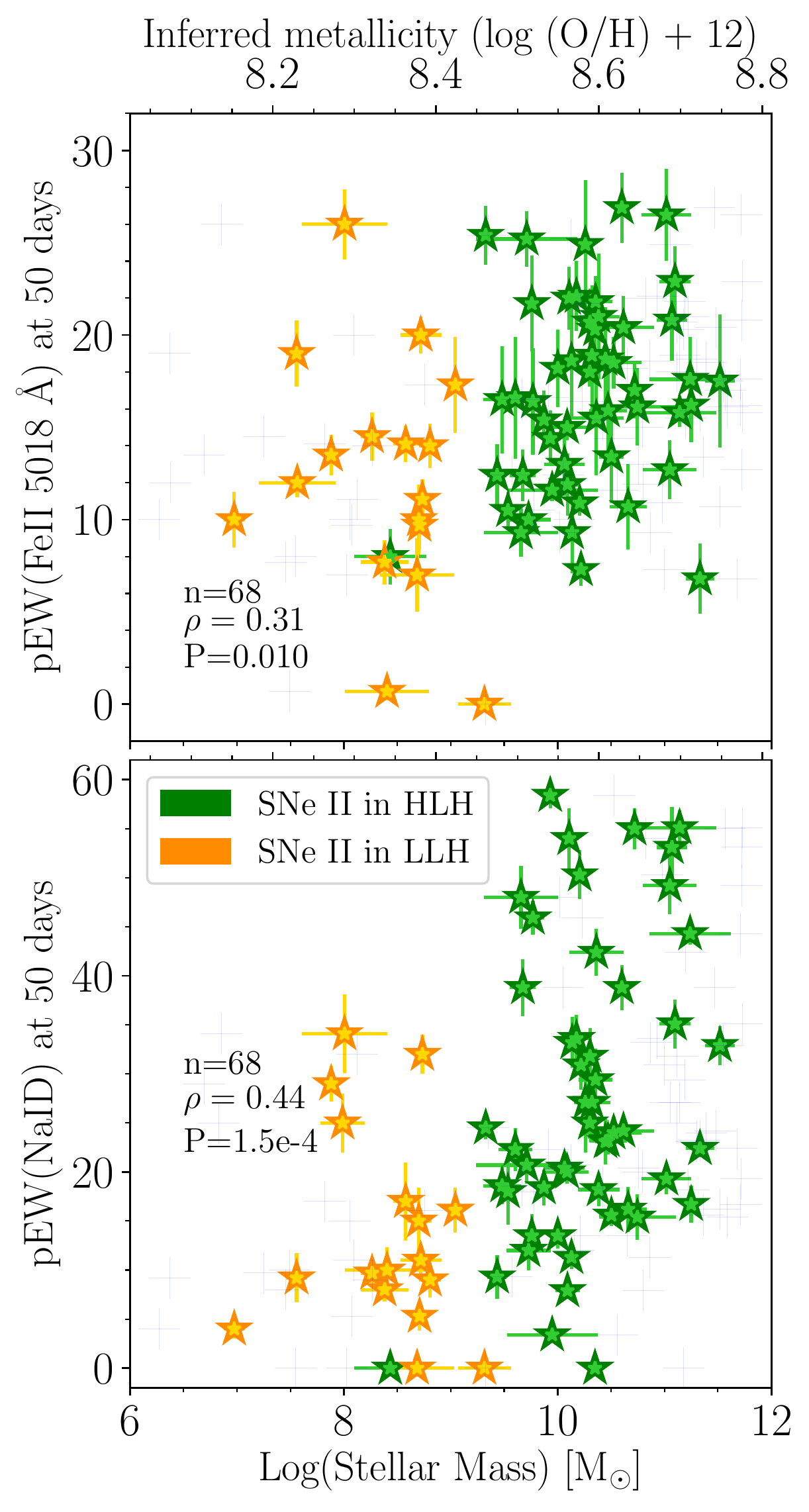}
\caption{The correlations between \pewfe\ and \mstellar\ (upper panel) and between  \pewNaD\ and \mstellar\ (lower panel). 
Green stars represent SNe~II in high-luminosity hosts, while yellow stars are SNe~II in low-luminosity hosts. In each panel,
$n$ is the number of events, $\rho$ is the Pearson correlation coefficient, and $P$ is the probability of detecting a 
correlation by chance. }
\label{fig:mstellar}
\end{figure}

Fig.~\ref{fig:mstellar} shows the relation between \pewfe\ and \pewNaD\ with \mstellar. The correlation between \mstellar\ and
\pewfe\ is weak ($\rho=0.31$), while the correlation between \mstellar\ and \pewNaD\ is moderate ($\rho=0.44$). SNe~II with 
a smaller \pewNaD\ preferentially occur in low \mstellar\ galaxies. By contrast, SNe~II with a higher \pewNaD\ are found 
in galaxies with a larger \mstellar. To test the significance of the correlation between \pewNaD\ and \mstellar, we use a
Monte Carlo resampling with $10^5$ random draws. Varying both parameters according to their 1-$\sigma$ uncertainties and a
Gaussian distribution, we estimate the Pearson coefficient ($\rho$) for each iteration. The median of $\rho$ is 0.47, which 
is consistent with our findings. 

Using a KS test, we probe the differences between SNe~II in low/high \mstellar. The KS test rejects the null hypothesis
that the distributions of the \pewfe\ in low-and high- \mstellar\ are drawn from populations with the same distribution.
The differences are more significant for \pewNaD. 
To test this result and using a Monte Carlo resampling (varying the pEW values using 1-$\sigma$ uncertainties according to 
a Gaussian distribution), we find a statistically significant difference between the \mstellar\ when comparing the \pewNaD. 

Given that our comparison models produce SN~II light curves and spectra that are qualitatively similar to observed SN~IIP (slow
decliners), we may expect stronger correlations when restricting our sample to slow-declining events. Surprisingly we find
the opposite: correlations decrease in strength when removing the fastest decliners.

\section{Discussion}
\label{discussion}

Using a sample of SNe~II and their host galaxies, we have examined various relationships between SN~II properties and their
host galaxies to probe the role of metallicity in SN~II evolution and diversity. Following \citetalias{Dessart14}, we tested
the influence of metallicity on the strength of the 
SN metal lines. Our main result is that SNe~II in lower-luminosity hosts display weaker metal lines (specifically \pewNaD, 
\pewfeb, \pewSc, \pewfe). Given the strong correlation between galaxy stellar mass (or luminosity) and galaxy metallicity,
this supports the potential for using SN~II spectral diagnostics as proxies for progenitor metallicity. In this section we 
compare our results with those found in previous studies to attempt to understand the role of galaxy properties in SN~II transient behaviour.

\subsection{Comparison to previous studies}

Using a sample of 119 events, \citetalias{Anderson16} found no evidence of a metallicity influence in the diversity of SNe~II. They examined 
potential correlations between metallicity and  various photometric parameters (\mmax, $s_{2}$ and \optd), but no significant trend was found.

Employing \pd\ as a proxy for the hydrogen envelope mass, instead of \optd, we similarly find no significant correlations. However, 
a KS test reject the null hypothesis that both $s_2$ samples (in high- and low-luminosity hosts) are drawn from parent populations
with the same distributions: slower declining SNe~II preferentially occur in low-luminosity galaxies. Unlike 
\citetalias{Taddia16}, we do not find a 
statistically significant difference between the samples when we compare their absolute magnitudes. 

Turning to the spectral parameters, \citetalias{Anderson16} found moderate correlations between the \pewfe\ and gas-phase metallicity as
measured through host \ion{H}{ii} region emission line analysis. Their results indicate that SNe~II with stronger iron lines explode in 
more metal-rich regions. As we do not have direct metallicity measurements of the SN host galaxies, we instead compare the \pewfe\ with 
the stellar mass of the host galaxy (derived from host photometry) and find a weaker correlation. When \citetalias{Anderson16} use their 
\lq gold\rq\ sample (i.e., SNe~IIP), their correlations increased. By contrast, in our plateau sample (i.e., SNe~II with
$s_{2}<1.5$\,mag\,100d$^{-1}$), the correlations are weaker.

The differences between these two studies are most likely due to the use, in the current study, of integrated stellar mass as a proxy for metallicity,
while \citetalias{Anderson16} derived metallicities of host \ion{H}{ii} regions directly through emission-line spectroscopy. Obtaining such
data for our sample will be the focus of a future project.

\citetalias{Dessart14} further found that, in their models, higher-metallicity SNe display weaker H$\alpha$ absorption.
Even though the differences in these values are small, they show an opposite behaviour to that found in this
work, where SNe~II in low-luminosity (low metallicity) hosts present a lower \pewHa. According to \citetalias{Dessart14}, the differences 
in the \pewHa\ do not originate from metallicity, but are instead probably more sensitive to time dependent effects, density profile,
mixing, clumping.

\subsection{Populating the metal-poor region with SNe II}

\citet{Stoll13}, \citetalias{Dessart14}, and more recently \citetalias{Anderson16} found a lack of SNe~II in low-metallicity environments.
Although this gap is expected because of an observational bias, some results have shown that SNe~II in lower-metallicity environments 
present smaller pEWs of the iron lines and no significant differences in their \pd\ \citepalias{Anderson16}.

\citetalias{Taddia16} showed how the observational gap found in the SN~II models at different metallicities was partially 
filled with their iPTF sample. However, when they compared the \pewfe\ with the metallicity of the host galaxy, they found a 
correlation with a weak significance, which could suggest that the observed lack of SNe~II is still visible.

\begin{figure}
\centering
\includegraphics[width=\columnwidth]{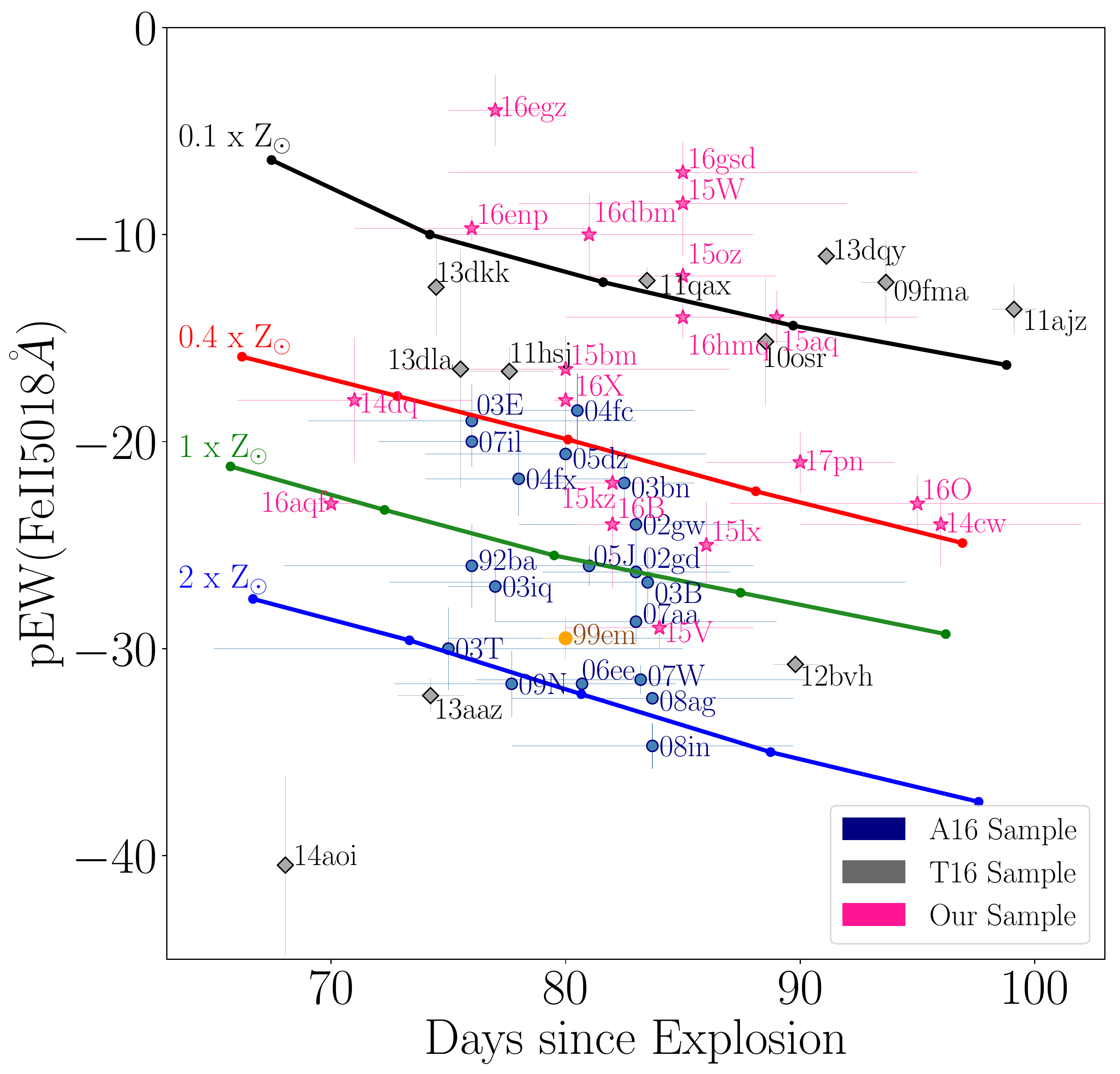}
\caption{Comparison of the \pewfe\ at around 80 days post-explosion between observed and model SN~II explosions. In light blue is
shown the CSP SNe~II with $s_2<1.5$ \citepalias{Anderson16}, in grey the (i)PTF sample \citepalias{Taddia16} and in magenta our 
SNe~II. The solid lines represent the \pewfe\ for four different models: 0.1Z$_\odot$ in black, 0.4Z$_\odot$ in red, Z$_\odot$ 
in dark green and 2Z$_\odot$ in blue.}
\label{fig:nube}
\end{figure}

Our sample probes lower-luminosity hosts and therefore lower inferred metallicities (Fig.~\ref{fig:mstellar}). These lower metallicities
are also evident when we compare our spectral measurements with the SN models at different metallicities. Reproducing Figure 6 of 
\citetalias{Dessart14} (the distribution of \pewfe\ around 80 days from explosion), 
it is possible to see how our sample fills the region between 0.1 and 0.4\,Z$_\odot$ (Fig.~\ref{fig:nube}). This figure 
includes the \citetalias{Anderson16} sample, the \citetalias{Taddia16} dataset and our SNe~II (magenta). Following
the process employed by \citetalias{Dessart14}, we measure the pEW for SNe~II with available spectra between 65 and 100 days 
post-explosion. A total of 19 SNe~II are incorporated in the pEW distribution plot. From \citetalias{Taddia16}, we also select SNe~II
within this range of time. Fig.~\ref{fig:nube} shows a significant number of SNe~II are consistent with progenitors with a metallicity
lower than 0.1\,Z$_\odot$. 

\subsection{Na I D as metallicity indicator?}

Recent studies (e.g., \citetalias{Anderson16}, \citetalias{Taddia16}) have shown the effects of environmental
metallicity in the strength of the iron lines in SNe~II spectra. However, these studies did not examine the potential influence of
metallicity on the strength of other metal lines, such as \ion{Sc}{ii}, \ion{Ba}{ii}, \ion{Na}{i} D. 

We test the relations between host-galaxy properties and the \pewHa\ absorption component, \pewHb, \pewNaD, 
\pewfe, \pewfeb, \pewBa\ and \pewSc. We find that, with the exception of \ion{Na}{i}\;D, these lines have only weak correlations with
host galaxy properties (see Table~\ref{table_pearson}). The \ion{Na}{i}\;D line shows a significant correlation with host galaxy 
luminosity and \mstellar. Fig.~\ref{fig:ewhost} and Fig.~\ref{fig:mstellar} present these results. Additionally, using the KS test,
we also find that \ion{Na}{i} D display statistically 
significant differences between the low- and high- luminosity groups.

This analysis suggests that \ion{Na}{i} D may be a good indicator of global properties of the galaxies, such as metallicity. This is 
surprising as, given the lightness of sodium, significant contamination of the hydrogen envelope with sodium created
from nuclear burning during the star's life is expected. These results suggest that either the mixing of such material to the outer envelope 
is very low, or that the degree of mixing also depends on progenitor metallicity, with higher metallicity progenitors undergoing 
more vigorous mixing during their lifetimes. However, a theoretical study on the production of sodium in massive stars at difference 
metallicity is needed in order to understand these results.

\section{Conclusions}
\label{conclusions}

In this work we have presented an analysis of SNe~II in low-luminosity galaxies, and compared to those SNe in high-luminosity hosts.
A total of 138 SNe~II were analysed using spectral diagnostics (velocities and pEWs) and photometric (\mmax, s$_2$ and \pd) properties, 
and compared with their host galaxy absolute magnitudes and stellar masses, which we use as a proxy for galaxy metallicity. Our main results are:

\begin{itemize}

\item We find that SNe~II in more metal-rich environments (i.e., a large stellar mass) display stronger metal lines in their photospheric
spectra. This is in agreement with the prediction from the models of \citetalias{Dessart14}. We found stronger correlations with \pewNaD\ than with \pewfe.

\item There is a commonly held view that the degree of hydrogen-envelope mass-stripping is
strongly dependent on progenitor metallicity in single star evolution. We do not
detect any signature of this, but it is likely that our RSG progenitors are drawn
mostly from the 8 - 15M${_\odot}$ range due to IMF statistics. At these masses and
luminosities, the mass-loss rates on the main sequence are not high enough
to have significant effect on the plateau ($\sim1-2$ M${_\odot}$ lost; \citealt{Eldridge04}), which is consistent with our measurements.

\item We find that all absorption lines in the spectra of SNe~II located in low-luminosity galaxies have smaller pEWs then those SNe~II 
in high-luminosity hosts. Comparing these results with models at different metallicities, we found that the H$\alpha$ absorption feature
shows the opposite behaviour. We found weak absorption in SNe~II in low luminosity hosts, while \citetalias{Dessart14} found stronger 
H$_{\alpha}$ absorption for low metallicity models. 

\item We find no evidence that expansion velocities (and therefore explosion energy) are affected by metallicity of the host. This suggests
there is physically little difference between the progenitor structure and explosions at low- and high-metallicity for RSGs that explode as SNe~II.   

\end{itemize}

\section*{Acknowledgements}

This work is based (in part) on observations collected at the European Organisation for Astronomical Research in the
Southern Hemisphere, Chile as part of PESSTO, (the Public ESO Spectroscopic Survey for Transient Objects Survey)
ESO program 188.D-3003, 191.D-0935, 191.D-0935, 197.D-1075, 199.D-0143. 
This work makes use of data from Las Cumbres Observatory, the Supernova Key Project, and the Global Supernova Project. 
Part of the funding for GROND (both hardware as well as personnel) was generously granted from the Leibniz-Prize to Prof. G.
Hasinger (DFG grant HA 1850/28-1). 

We thank the annonymous referee for the useful suggestions.
We are grateful to Andrea Pastorello, Jos\'e Luis Prieto, Avet Harutyunyan and R. Mark Wagner for performing some of the observations used in this work.

C.P.G. and M.S. acknowledge support from EU/FP7-ERC grant No. [615929]. L.G. was supported in part by the US National Science Foundation under Grant AST-1311862.
Support for I.A. was provided by NASA through the Einstein Fellowship Program, grant PF6-170148. 
T.-W.Chen acknowledgments the funding provided by the Alexander von Humboldt Foundation.
D.A.H., G.H., and C.M. are supported by NSF grant AST-1313484.
G.L. was supported by a research grant (19054) from VILLUM FONDEN.
M.G. is supported by the Polish National Science Centre grant OPUS 2015/17/B/ST9/03167. 
Support for F.O.E. is provided by FONDECYT through grant 11170953 and by the Ministry of Economy, Development, and Tourism's Millennium 
Science Initiative through grant IC120009, awarded to The Millennium Institute of Astrophysics, MAS.
Support for G.P. is provided by the Ministry of Economy, Development, and Tourism's Millennium Science Initiative through 
grant IC120009, awarded to The Millennium Institute of Astrophysics, MAS.
S.J.S. acknowledges funding from ERC Grant 291222  and STFC grant Grant Ref: ST/P000312/1 

This research has made use of the NASA/IPAC Extragalactic Database (NED) which is operated by the Jet Propulsion Laboratory,
California Institute of Technology, under contract with the National Aeronautics. We acknowledge the usage of the HyperLeda 
database (http://leda.univ-lyon1.fr)

Funding for the Sloan Digital Sky Survey IV has been provided by the Alfred P. Sloan Foundation, the U.S.
Department of Energy Office of Science, and the Participating Institutions. SDSS-IV acknowledges support 
and resources from the Center for High-Performance Computing at the University of Utah. The SDSS web site 
is www.sdss.org. SDSS-IV is managed by the Astrophysical Research Consortium for the Participating 
Institutions of the SDSS Collaboration including the Brazilian Participation Group, the Carnegie Institution 
for Science, Carnegie Mellon University, the Chilean Participation Group, the French Participation Group, 
Harvard-Smithsonian Center for Astrophysics, Instituto de Astrof\'isica de Canarias, The Johns Hopkins University, 
Kavli Institute for the Physics and Mathematics of the Universe (IPMU) / University of Tokyo, Lawrence Berkeley 
National Laboratory, Leibniz Institut f\"ur Astrophysik Potsdam (AIP), Max-Planck-Institut f\"ur Astronomie 
(MPIA Heidelberg), Max-Planck-Institut f\"ur Astrophysik (MPA Garching), Max-Planck-Institut f\"ur Extraterrestrische
Physik (MPE), National Astronomical Observatories of China, New Mexico State University, New York University, 
University of Notre Dame, Observat\'ario Nacional / MCTI, The Ohio State University, Pennsylvania State University, 
Shanghai Astronomical Observatory, United Kingdom Participation Group, Universidad Nacional Aut\'onoma de M\'exico,
University of Arizona, University of Colorado Boulder, University of Oxford, University of Portsmouth, 
University of Utah, University of Virginia, University of Washington, University of Wisconsin, 
Vanderbilt University, and Yale University.

The Pan-STARRS1 Surveys (PS1) have been made possible through contributions of the Institute for Astronomy,
the University of Hawaii, the Pan-STARRS Project Office, the Max-Planck Society and its participating 
institutes, the Max Planck Institute for Astronomy, Heidelberg and the Max Planck Institute for Extraterrestrial 
Physics, Garching, The Johns Hopkins University, Durham University, the University of Edinburgh, Queen's 
University Belfast, the Harvard-Smithsonian Center for Astrophysics, the Las Cumbres Observatory Global Telescope
Network Incorporated, the National Central University of Taiwan, the Space Telescope Science Institute, the
National Aeronautics and Space Administration under Grant No. NNX08AR22G issued through the Planetary Science
Division of the NASA Science Mission Directorate, the National Science Foundation under Grant No. AST-1238877,
the University of Maryland, and Eotvos Lorand University (ELTE).





\clearpage
{\onecolumn
\begin{landscape}
\input{table_info.tex} 
\clearpage
\end{landscape}
}

\clearpage
\begin{landscape}
\input{table_obs.tex}
\clearpage
\end{landscape}

\clearpage
{\onecolumn
\begin{landscape}
\input{table_mag.tex}
\clearpage
\end{landscape}
}

\clearpage
{\onecolumn
\begin{landscape}
\input{table_measure.tex}
\clearpage
\end{landscape}
}

\input{table_summary.tex}

\input{table_values.tex}

\input{table_pearson.tex}

\label{lastpage}
\end{document}

%% file: table_info.tex
\scriptsize
\centering                                                                                                                                       
\begin{longtable}[t]{lccccccclccccc}                                                                                                             
\caption{SN~II sample}                                                                                                                           
\label{table_info}\\                                                                                                                             
\hline                                                                                                                                           
\hline                                                                                                                                           
                  &              Host            & Recession     &   Host M$_{B}$    & $E(B-V)_{\rm MW}$  & Discovery &    Source                & Discovery  &    Explosion       \\
       SN         &             Galaxy     & velocity (km s$^{-1}$) & type           &   (mag)            & date MJD  &    discovery             &  Reference &       Epoch         \\
\hline                                                                                                                                           
\hline                                                                                                                                           
SN2009lq	  &     Anonymous                &  13290$^{1}$  &  \nodata          &  0.027             & 55158.5   &     CHASE                & CBET 2039  & 55156.1$^{n}$(2.5) \\[0.3ex]
ASASSN-14dq       &     UGC 11860                &  3125         &  $-18.30$         &  0.062             & 56847.0   &    ASAS-SN               & ATEL 6301  & 56841.0$^{n}$(5)   \\[0.3ex]
SN2014cw	  & 	PGC 68414                &  1800         &  \nodata          &  0.052             & 56899.0   &    MASTER                & ATEL 6435  & 56892.6$^{s}$(6)   \\
\hline
\hline
\end{longtable}
\begin{list}{}{}
\item SNe and host galaxy information. Columns: (1) SN name; (2) galaxy  name; (3) the host galaxy heliocentric recession velocity. These are taken
from the Nasa Extragalactic Database (NED: http://ned.ipac.caltech.edu/) unless indicated by a superscript (sources in table notes); 
(4) host galaxy absolute $B-$band magnitudes (taken from the LEDA database: http://leda.univ-lyon1.fr/); (5) the reddening due to dust in our Galaxy
\citep{Schlafly11} taken from NED; (6) discovery date; (7) source of discovery: CHilean Automatic Supernovas sEarch (CHASE; http://www.das.uchile.cl/proyectoCHASE/), 
All-Sky Automated Survey for Supernovae (ASAS-SN; http://www.astronomy.ohio-state.edu/), MASTER Global Robotic Net (http://observ.pereplet.ru/), 
Lick Observatory Supernova Search (LOSS; http://w.astro.berkeley.edu/bait/kait.html), Catalina Real-Time Transient Survey (CRTS; http://nesssi.cacr.caltech.edu/CRTS/), 
Optical Gravitational Lensing Experiment (OGLE-IV), Transient Detection System (http://ogle.astrouw.edu.pl/ogle4/transients/2015/transients.html), 
Pan-STARRS1 (https://panstarrs.stsci.edu/), Gaia Photometric Science Alerts (GaiaAlerts, http://gsaweb.ast.cam.ac.uk/alerts/), ATLAS (http://fallingstar.com/home.php);
Perth Astronomical Research Group (PARG; http://www.parg.asn.au/); Nearby Supernova Factory (SNFactory; https://snfactory.lbl.gov/);
Alain Klotz (http://alain.klotz.free.fr/snalert/); Takao Doi (http://iss.jaxa.jp/en/astro/biographies/doi/index.html);
Sloan Digital Sky Survey Collaboration (SDSS; http://www.sdss.org/collaboration/); Lick Observatory and Tenagra Observatory Supernova Search (LOTOSS; http://w.astro.berkeley.edu/bait/lotoss.html); 
Berto Monard (http://assa.saao.ac.za/about/awards/gill-medal/berto-monard-awarded-2004-gill-medal/); Tom Boles (http://www.coddenhamobservatories.org/);
Lulin Observatory (http://www.lulin.ncu.edu.tw/); CEAMIG/REA Supernovae Search (http://www.ceamig.org.br/); Tenagra Observatory (http://www.tenagraobservatories.com/);
Brazilian Supernovae Search (BRASS; http://brass.astrodatabase.net/index.htm); Perth Observatory (https://www.wa.gov.au/perthobs/); Ron Arbour (http://mstecker.com/pages/apparbour.htm);
Lunar and Planetary Laboratory (LPL; https://www.lpl.arizona.edu/); Col Drusci\'e Remore Observatory Supernovae Search (CROSS; http://www.cortinastelle.it/snindex.htm); 
(8) discovery reference; (9) explosion epoch. 
They are estimated using the SN non-detection or
through the spectral matching. More details can be found in \citet{Gutierrez17a}.\\
\item $^{1}$ Redshift obtained from the SN spectrum.\\
\item $^{2}$ Taken from the Asiago supernova catalog: http://graspa.oapd.inaf.it/ \citep{Barbon99}.\\
\item $^{\star}$ Discovery epoch.\\
\item $^{s}$ Explosion epoch estimation through spectral matching. \\
\item $^{n}$ Explosion epoch estimation from SN non-detection. \\
\end{list}

%% file: table_obs.tex
\small
\begin{table}
\centering
\small
\caption{Observation details}
\label{obs}
\begin{tabular}[t]{ccccccccccc}
\hline
SN 		  &  Photometry	& Filters	& Photometry 	         &   \#    & Spectroscopic$^{\ast}$  & Spectra       & \#      & Source of Spectra  \\
		  &   source	&		& coverage (d)	         & Points  &    source               & coverage (d)  & Spectra & around 50 days     \\
\hline                                                                                                       
\hline                                                                                                       
SN2009lq	  & PROMPT1,3,5	& $BVRI$	& 140			 & 17	   & LDSS3+WFCCD             &  120	     &  2      &    \nodata         \\
ASASSN-14dq       & LCO         & $BVgri$       & 290$^{\star}$ 	 & 40      & FLOYDS                  &  96           &  5      &    \nodata         \\
SN2014cw	  & PROMPT1,3,5	& $BVRI$        & 350         		 & 8       & EFOSC2+LRS+GS+OSMOS     &  114          &  4      &    GS              \\
ASASSN-14kp       & PROMPT1,3,5	& $BVRI$        & 35            	 & 18      & EFOSC2+WFCCD            &  76           &  10     &    EFOSC2          \\
SN2015V           & LCO         & $BVgri$       & 210           	 & 28      & FLOYDS                  &  215          &  13     &    FLOYDS          \\
SN2015W           & LCO         & $BVgri$       & 120       	 	 & 58      & FLOYDS                  &  96           &  13     &    FLOYDS          \\
SN2015aq          & LCO         & $BVgri$       & 390$^{\clubsuit}$      & 42      & FLOYDS                  &  276          &  15     &    FLOYDS          \\
SN2015bm          & LCO         & $gri$         & 140                    & 30      & EFOSC2                  &  109          &  11     &    EFOSC2          \\
SN2015bs          & CSP II      & $BVri$        & 80                     & 16      & EFOSC2                  &  81           &  8      &    EFOSC2          \\   
ASASSN-15kz       & LCO         & $BVgri$       & 90                     & 30      & FLOYDS                  &  82           &  8      &    FLOYDS          \\
ASASSN-15lx       & LCO         & $BVgri$       & 340$^{\bigtriangleup}$ & 62      & FLOYDS                  &  150          &  17     &    FLOYDS          \\
ASASSN-15oz       & LCO         & $BVgri$       & 400$^{\diamond}$       & 73      & EFOSC2+FLOYDS           &  390          &  21     &    FLOYDS          \\
ASASSN-15rp       & \nodata     & \nodata       & \nodata                & \nodata & EFOSC2                  &  77           &  3      &    EFOSC2          \\
SN2016B           & LCO         & $BVgri$       & 510$^{\triangleleft}$  & 63      & EFOSC2+FLOYDS           &  390          &  30     &    EFOSC2          \\
SN2016O           & LCO         & $gri$         & 110                    & 14      & EFOSC2                  &  94           &  8      &    EFOSC2          \\
SN2016X           & LCO         & $BVgri$       & 190  		         & 46      & EFOSC2+FLOYDS           &  141          &  29     &    EFOSC2          \\
SN2016aqf         & LCO         & $BVgri$       & 300$^{\triangleright}$ & 33      & EFOSC2+FLOYDS           &  326          &  29     &    EFOSC2          \\
SN2016ase         & LCO         & $BVgri$       & 150                    & 45      & EFOSC2+FLOYDS           &  147          &  16     &    FLOYDS          \\
SN2016blz         & LCO         & $BVgri$       & 150                    & 34      & FLOYDS                  &  143          &  14     &    FLOYDS          \\
SN2016dbm         & LCO         & $BVgri$       & 110                    & 25      & EFOSC2                  &  107          &  18     &    EFOSC2          \\
SN2016dpd         & LCO         & $BVgri$       & 130                    & 23      & FLOYDS                  &  53           &  8      &    FLOYDS          \\
SN2016drl         & LCO         & $gri$         & 80                     & 10      & EFOSC2                  &  59           &  7      &    EFOSC2          \\
SN2016egz         & LCO         & $BVgri$       & 300$^{\bullet}$        & 50      & EFOSC2+FLOYDS           &  179          &  34     &    EFOSC2          \\
SN2016enk         & LCO         & $BVgri$       &257$^{\bigtriangledown}$& 28      & FLOYDS                  &  190          &  8      &    FLOYDS          \\
SN2016enp         & LCO         & $gri$         & 150                    & 22      & EFOSC2                  &  104          &  11     &    EFOSC2          \\
SN2016gsd         & LCO         & $BVgri$       & 90                     & 20      & EFOSC2+FLOYDS           &  110          &  9      &    EFOSC2          \\
SN2016hmq         & LCO         & $BVgri$       & 110                    & 18	   & EFOSC2                  &  100          &  8      &    EFOSC2          \\
SN2016hpt         & LCO         & $BVgri$       & 40                     & 6       & EFOSC2                  &  59           &  3      &    EFOSC2          \\
SN2017pn          & LCO         & $BVgri$       & 70                     & 13      & EFOSC2                  &  91           &  5      &    EFOSC2          \\
SN2017vp          & GROND	& $g'r'i'z'JHK$ & 110                    & 12      & EFOSC2                  &  50           &  4      &    EFOSC2          \\
\hline
\hline
\end{tabular}
\begin{list}{}{}
\item \textbf{Notes:} \textbf{LDSS3:} Low Dispersion Survey Spectrograph at the Magellan Clay 6.5-m telescope; 
\textbf{WFCCD}: Wide Field CCD Camera at the 2.5-m du Pont Telescope;
\textbf{FLOYDS:} FLOYDS spectrographs on the Faulkes Telescope South (FTS) and the Faulkes Telescope North (FTN);
\textbf{EFOSC2:} ESO Faint Object Spectrograph and Camera at the 3.5-m ESO New Technology Telescope (NTT);
\textbf{LRS:} Low Resolution Spectrograph at the 3.6m Telescopio Nazionale Galileo (TNG);
\textbf{GS:} Goodman Spectrograph at the SOAR 4.1-m telescope; 
\textbf{OSMOS:} Ohio State Multi-Object Spectrograph at the 2.4-m Hiltner Telescope.
\item Note that ASASSN-15rp does not have Photometric information.\\
$^{\star}$ Break in the observations for $\sim140$ days.  \\
$^{\clubsuit}$ Break in the observations  for $\sim150$ days. \\[0.3ex]
$^{\bigtriangleup}$ Break in the observations n for $\sim150$ days. \\[0.3ex]
$^{\diamond}$ Break in the observations  for $\sim100$ days. \\[0.3ex]
$^{\triangleleft}$ Break in the observations for $\sim130$ days. \\[0.3ex]
$^{\triangleright}$ Break in the observations for $\sim70$ days. \\[0.3ex]
$^{\bullet}$ Break in the observations for $\sim100$ days. \\[0.3ex]
$^{\bigtriangledown}$ Break in the observations for $\sim69$ days. \\[0.3ex]
\end{list}
\end{table}

%% file: table_mag.tex
\clearpage
\centering
\small
\begin{longtable}[t!]{lcccccccccccccccc}
\caption{Host galaxy information}
\label{mag} \\
\hline
SN 		      &     Host M$_{B}$   &    Host M$_{u}$   &   Host M$_{g}$    &    Host M$_{r}$     &     Host M$_{i}$    &    Host M$_{z}$      &  Log(\mstellar)   \\	
		      &        (mag)       &	      (mag)    &      (mag)	   &      (mag)          &        (mag)	       &        (mag)         &    (M$_\odot$)    \\
\hline                                                                                                                  
\hline                                 
SN2009lq              &  \nodata	   &  \nodata          &  \nodata          &   $>-14.39\pm0.10$  &   \nodata           &   \nodata            &  \nodata          \\  
ASASSN-14dq           &  $-18.30\pm0.50$   &  \nodata          &  $-16.99\pm0.01$  &   $-17.28\pm0.01$   &   $-17.37\pm0.01$   &   $-17.60\pm0.03$    &  $8.72\pm0.19$    \\   
SN2014cw              &  \nodata	   &  \nodata          &  \nodata          &   $-13.57\pm0.04$   &   $-13.45\pm0.01$   &   $-12.78\pm0.08$    &  $6.97\pm0.10$    \\
ASASSN-14kp           &  $-16.87\pm0.52$   &  \nodata          &  \nodata          &   \nodata	         &   \nodata           &   \nodata            &  \nodata          \\
SN2015V               &  $-18.50\pm0.36$   &  \nodata          &  $-17.45\pm0.01$  &   $-17.76\pm0.02$   &   $-17.67\pm0.03$   &   $-18.07\pm0.01$    &  $8.72\pm0.19$    \\
\hline
\hline
\end{longtable}
\begin{list}{}{}
\item $^{\bullet}$ SNe~II from \citetalias{Anderson16} included in the low-luminosity group.
\item $^{\diamond}$ SNe~II from our sample included in the high-luminosity group.
\end{list}

%% file: table_measure.tex
\scriptsize
\centering
\setlength{\tabcolsep}{5pt}
\begin{longtable}[t]{lccccccccccccc}
\caption{Photometric and spectroscopic (+50 days) measurements}
\label{table_measure}\\
\hline
SN 	          &      \mmax\       &     s$_2$        &     \pd\       &    \optd\       &    \pewHa\      &    \pewHb\     &    \pewfe\     &    \pewfeb\    & \pewNaD\       &    \pewBa\    &    \pewSc\    & vel(\ion{Na}{i}\,D) \\[0.3ex]
                  &      (mag)        & (mag 100d$^{-1}$)&     (days)     &    (days)       &      (\AA)      &     (\AA)      &      (\AA)     &      (\AA)     &    (\AA)       &      (\AA)    &     (\AA)     &    (km.s$^{-1}$)    \\[0.3ex]
\hline                                                                                    
\hline                                                                                    
SN2009lq	  &  $-17.44\pm0.10$  &  $0.27\pm0.05$   &  $33.7\pm2.1$  &  $75.8\pm2.9$   &   \nodata       &  \nodata	&  \nodata	 &  \nodata       &  \nodata       &  \nodata      &  \nodata      &  \nodata	      \\[0.3ex]
ASASSN-14dq       &  $-16.70\pm0.10$  &  $1.10\pm0.10$   &  \nodata	  &  $88.6\pm0.7$   &   \nodata       &  \nodata	&  \nodata	 &  \nodata       &  \nodata       &  \nodata      &  \nodata      &  \nodata	      \\[0.3ex]
SN2014cw	  &  \nodata          &  \nodata	 &  \nodata	  &  \nodata	    &   $60.0\pm3.6$  &  $62.0\pm3.2$  &  $10.1\pm1.5$  &  $19.1\pm1.4$  &  $4.1\pm1.1$   &  $0.0\pm0.0$  &  $0.0\pm0.0$  &  $3220\pm663$     \\[0.3ex]
ASASSN-14kp       &  \nodata          &  \nodata	 &  \nodata	  &  \nodata	    &   \nodata       &  $45.5\pm4.3$  &  $13.0\pm1.5$  &  $27.0\pm4.5$  &  $0.0\pm0.0$   &  $0.0\pm0.0$  &  $0.0\pm0.0$  &  \nodata	      \\[0.3ex]
SN2015V           &  $-15.74\pm0.10$  &  $0.31\pm0.01$   &  $68.6\pm1.2$  &  $96.4\pm1.1$   &   $80.0\pm2.0$  &  $54.0\pm3.0$  &  $20.2\pm1.0$  &  $30.2\pm2.0$  &  $11.0\pm1.5$  &  $3.0\pm0.9$  &  $2.8\pm0.8$  &  $3780\pm660$     \\
\hline
\hline
\end{longtable} 

%% file: table_summary.tex
\begin{table}
\centering
\small
\caption{Sample details}
\label{table_summary}
\begin{tabular}[t]{lcccccccccc}
\hline
                                 & SNe~II in LLH  &  SNe~II in HLH  &  Total        \\ 
\hline                                                              
\hline                                                              
Full Sample                      & \textbf{35}    &  \textbf{103}   &  \textbf{138} \\
\hline
New SNe                          &       27       &        3        &   30          \\
SNe from \citetalias{Anderson16} &       8        &       100       &  108          \\
\hline
SNe with light curve in $V-$band &      29        &       103       &  132          \\
SNe with spectra around 50 days  &      28        &        72       &  100          \\
SNe with \mbhost\ data           &      19        &        95       &  114          \\
SNe with \mrhost\ data           &      28        &        74       &  102           \\
\hline
\hline
\end{tabular}    
\end{table}

%% file: table_values.tex
\begin{table}
\centering
\small
\caption{Mean values and KS test results}
\label{table_values}
\begin{tabular}[t]{lcccccccccc}
\hline
Parameter                  &           Mean values in the        &      Mean values in the             & $\rho_\mathrm{KS}$            \\[0.3ex]	
                           &         high-luminosity             &         low-luminosity              &                               \\[0.3ex]	
		           &            host group               &          host group                 &                               \\[0.3ex]
\hline                                                                                        
\hline                     
		           &                                     &                                     &                               \\[0.3ex]
\mmax\                     &  $-16.67\pm0.10$ mag (93)           &  $-17.06\pm0.22$ mag (26)           &     0.174                     \\[0.3ex] 
s$_2$                      &  $1.19\pm0.09$ mag 100d${-1}$ (96)  &  $0.72\pm0.13$ mag 100d$^{-1}$ (25) & \textbf{9.63$\times10^{-4}$}  \\[0.3ex]  
\pd\                       &  $49.88\pm2.16$ days (28)           &  $46.69\pm3.89$ days (14)           &     0.369                     \\[0.3ex]
\optd\			   &  $83.98\pm2.03$ days (59)           &  $86.62\pm3.72$ days (19)           &     0.769                     \\
\hline
\hline
\end{tabular}
\begin{list}{}{}
\item The number of events used for each measurement are given in parenthesis.
\item $\rho_\mathrm{KS}$ lower than 0.01 are presented in bold.
\end{list}         
\end{table}

%% file: table_pearson.tex
\begin{table}
\centering
\scriptsize
\caption{Pearson correlation coefficients}
\label{table_pearson}
\begin{tabular}[t]{lcccccccccc}
\hline
Parameter                 &  Host M$_{B}$                     &  Host M$_{r}$                     &  Host M$_{i}$                     &  Host M$_{z}$                     &  \mstellar                       \\[0.3ex]	
                          &                                   &                                   &                                   &                                   &                                  \\[0.3ex]	
\hline                                                                                                                                                                    
\hline                                                                                                                                                                   
                          &                                   &                                   &                                   &                                   &                                   \\
\mmax\                    &  $-0.04$ (101; $0.73$)            &  $-0.22$ (88; $0.04$)             &  $-0.14$ (85; $0.19$)             &  $-0.12$ (82; $0.29$)             &  $0.19$ (82; $0.08$)             \\[0.3ex]
s$_2$                     &  $-0.19$ (104; $0.06$)            &  $-0.22$ (90; $0.04$)             &  $-0.20$ (87; $0.07$)	      &  $-0.17$ (84; $0.11$)             &  $0.18$ (84; $0.10$)             \\[0.3ex]
\pd\                      &  $-0.04$ (38;  $0.81$)            &  $-0.19$ (28; $0.32$)             &  $-0.04$ (26; $0.84$)             &  $0.22$  (24; $0.30$)             &  $-0.13$ (25; $0.53$)            \\[0.3ex]
\optd\                    &  $0.02$ (67;  $0.85$)             &  $-0.11$ (56; $0.44$)             &  $0.03$ (54; $0.86$)              &  $0.09$  (52; $0.52$)             &  $0.22$ (59; $0.09$)            \\
\hline
\hline
\end{tabular}
\begin{list}{}{}
\item The number of events and the probability of finding such correlation by change are presented in parenthesis.
\end{list}         
\end{table}